\documentclass[]{interact}

\usepackage{epstopdf}

\usepackage[nolist]{acronym}
\usepackage{graphicx}
\usepackage{subcaption}
\usepackage{wrapfig}
\usepackage[hidelinks]{hyperref}
\usepackage{color}
\usepackage{rotating}
\usepackage{xurl}

\usepackage[natbibapa,nodoi]{apacite}
\theoremstyle{plain}

\theoremstyle{definition}

\theoremstyle{remark}

\begin{document}

\begin{acronym}[]
	\acro{AI}{Artificial Intelligence}
	\acro{ACC}{Adaptive Cruise Control}
	\acro{ADAS}{Advanced Driver Assistance System}
	\acro{CAN}{Controller Area Network}
	\acro{ECU}{Electronic Control Unit}
	\acro{GDPR}{General Data Protection Regulation}
	\acro{GOMS}{Goals, Operators, Methods, Selection rules}
	\acro{HCI}{Human-Computer Interaction}
	\acro{HMI}{Human-Machine Interaction}
	\acro{HU}{Head Unit}
	\acro{IVIS}{In-Vehicle Information System}
	\acro{KLM}{Keystroke-Level Model}
	\acro{KPI}{Key Performance Indicator}
	\acro{LKA}{Lane Keeping Assist}
	\acro{LCA}{Lane Centering Assist}
	\acro{MGD}{Mean Glance Duration}
	\acro{OEM}{Original Equipment Manufacturer}
	\acro{OTA}{Over-The-Air}
	\acro{AOI}{Area of Interest}
	\acro{RF}{Random Forest}
	\acro{SA}{Steering Assist}
	\acro{SHAP}{SHapley Additive exPlanation}
	\acro{TGD}{Total Glance Duration}
	\acro{UCD}{User-centered Design}
	\acro{UX}{User Experience}
\end{acronym}


\title{Multitasking while Driving: How Drivers Self-Regulate their Interaction with In-Vehicle Touchscreens in Automated Driving}

\author{
\name{Patrick Ebel\textsuperscript{a}\thanks{This is the author’s version of the work. It is posted here for your personal use. Not for redistribution. The definitive Version of Record was published in the International Journal of Human–Computer Interaction. CONTACT Patrick Ebel. Email: ebel@cs.uni-koeln.de}, Christoph Lingenfelder\textsuperscript{b} and Andreas Vogelsang\textsuperscript{a}}
\affil{\textsuperscript{a}University of Cologne, Cologne, Germany; \textsuperscript{b} MBition GmbH, Berlin, Germany}
}

\maketitle

\begin{abstract}
Driver assistance systems are designed to increase comfort and safety by automating parts of the driving task. At the same time, modern in-vehicle information systems with large touchscreens provide the driver with numerous options for entertainment, information, or communication, and are a potential source of distraction. However, little is known about how driving automation affects how drivers interact with the center stack touchscreen, i.e., how drivers self-regulate their behavior in response to different levels of driving automation. To investigate this, we apply multilevel models to a real-world driving dataset consisting of 31,378 sequences. Our results show significant differences in drivers' interaction and glance behavior in response to different levels of driving automation, vehicle speed, and road curvature. During automated driving, drivers perform more interactions per touchscreen sequence and increase the time spent looking at the center stack touchscreen. Specifically, at higher levels of driving automation (level 2), the mean glance duration toward the center stack touchscreen increases by 36\,\% and the mean number of interactions per sequence increases by 17\,\% compared to manual driving. Furthermore, partially automated driving has a strong impact on the use of more complex UI elements (e.g., maps) and touch gestures (e.g., multitouch). We also show that the effect of driving automation on drivers' self-regulation is greater than that of vehicle speed and road curvature. The derived knowledge can inform the design and evaluation of touch-based infotainment systems and the development of context-aware driver monitoring systems.
\end{abstract}


\section{Introduction}

Driver distraction is one of the main causes of motor vehicle crashes. The primary goal of automated driving functions like \ac{ACC} and \ac{LCA}, apart from making driving more comfortable, is to make driving safer. Multiple studies show that these systems can make driving safer by an increased time headway and that they reduce the incidence of critical situations~\citep{Faber.2012, Ervin.2005}. However, even though automated driving functions are more widely available and powerful than ever, the number of crashes based on human error due to distraction stagnated in recent years~\citep{NHTSA.2021}. Studies show that driving automation does not only positively affect driving safety but also tends to increase the margins in which drivers consider it safe to engage in non-driving-related tasks~\citep{risteska.2021, morando.2019, dunn.2020}. To interact with \acp{IVIS} or mobile phones while driving, drivers need to distribute their attention between the primary driving task and the non-driving-related secondary task. Although drivers are proven to self-regulate their secondary task engagements based on driving demands~\citep{christoph.2019, onate-vega.2020, oviedo-trespalacios.2019}, this task-switching behavior is directly associated with an increased crash risk~\citep{dingus.2016}. This is particularly critical as drivers tend to overestimate the capabilities of automated driving functions~\citep{deguzman.2021} potentially making it more likely to engage in non-driving-related tasks~\citep{dunn.2020} in situations in which they are supposed to monitor these functions constantly~\citep{SAE.2021}.

As modern \acp{IVIS} continue to improve and large center stack touchscreens are becoming the main interface between driver and vehicle, the temptation for drivers to interact with them is likely to increase~\citep{starkey.2020}. A deep understanding of how drivers self-regulate their secondary task engagements in response to varying driving demands can facilitate the design of \acp{IVIS} that are safe to use in all situations~\citep{ebel.2021}. Knowing what naturally feels safe for drivers can also, improve attention management systems to provide situation-dependent interventions when inappropriate self-regulation is detected~\citep{risteska.2021}.
To better understand how drivers adapt their engagement in secondary touchscreen tasks, we investigate the effect of driving automation (manual vs. \ac{ACC} vs. \ac{ACC} + \ac{LCA}), vehicle speed, and road curvature on drivers' tactical and operational self-regulation. We further show how the effect of driving automation depends on vehicle speed and road curvature. Therefore, we employ multilevel modeling on a real-world driving dataset consisting of 31,378 user interaction sequences and the accompanying driving and eye tracking data. 

To evaluate tactical self-regulation, we fit generalized linear mixed models estimating the probability of drivers interacting with specific UI elements. Our results show that drivers self-regulate their interaction behavior differently across the UI elements. During ACC+LCA driving, the odds of a driver interacting with a map element are, for example, 1.62 times as high as for manual driving. The probability to interact with a regular button, however, remains similar.

Furthermore, we measure drivers' operational self-regulation as glance behavior adaptions. The multilevel modeling results indicate that drivers adapt their glance behavior based on automation level, vehicle speed, and road curvature. Across all driving situations, the mean glance duration increases by 12\,\% for ACC driving compared to manual driving and by 36\,\% for ACC+LCA driving. The odds that drivers perform a glance longer than 2 seconds are 1.6 and 3.6 times as high, respectively.

\subsection{Relation to Previous Publications}

This paper is an extension of a previous publication~\citep{ebel.2022}.
In the article at hand, we make the following additional contributions:

\begin{itemize}
    \item We base our analysis on three times as much data as in the previous paper. In our previous work, we analyzed 10,139 driving sequences from a time period of three months (Oct 2021--Feb 2022). In the paper at hand, we have extended this dataset to 31,378 driving sequences from one year (Oct 2021--Oct 2022).
    \item We have refined the data processing and the statistical models. First, in contrast to our previous work, we removed all sequences during which the driver did not perform a gaze transition between the center stack touchscreen and the road. Therefore, we only consider sequences during which regulations actually happened. Secondly, we extended our statistical models to also account for different car types as random effects. 
    \item In the previous paper, we only analyzed the effect of driving automation, vehicle speed, and road curvature on glance behavior and interactions with specific UI elements. We now also analyze how these factors affect the number of interactions and the number of touch gestures within a sequence.
\end{itemize}

Due to the improvements in data processing and the larger dataset, the results of this paper are not directly comparable with the results of our previous paper in terms of absolute numbers. However, our results confirm the general trends observed in the previous paper.  
In particular, this paper confirms the finding that, at the tactical level, drivers' self-regulation of complex touchscreen interactions is more sensitive to driving demand than that of simple interactions.

\subsection{Driver Distraction}\label{ch:DriverDistraction}

Driving a car is a complex task. It requires drivers to simultaneously perform different activities. They need to watch and follow the road, perform steering and pedal movements, and react to sudden changes in the driving environment~\citep{regan.2022}. Despite the complexity of the driving task, drivers tend to engage in non-driving related activities like talking to the passenger or interacting with the smartphone or the \ac{IVIS}. \citet{regan.2009} describe the interaction with devices like mobile phones or \acp{IVIS} as a competing activity. These interactions compete with the resources required to perform activities critical for safe driving. Thus, driver distraction is defined as the \textit{``diversion of attention away from activities critical for safe driving toward a competing activity''}~\citep{lee.2008}. Whereas various types of driver distraction exist, we will focus on \textit{visual distraction}. Visual distraction is concerned with drivers taking their eyes off the road.  Thus it is also described as the ``[d]iversion of attention towards things that we see''~\citep{regan.2022}. Studies show that visual distraction is correlated with increased crash risk. \citet{Klauer.2006} found that glances off the road longer than two seconds increase the crash risk by two times compared to normal driving. Accordingly, the ``Visual-Manual NHTSA Driver Distraction Guidelines for In-Vehicle Electronic Devices''~\citep{NHTSA.2014} define upper bounds for glances longer than two seconds. This shows that visual distraction is an important factor that needs to be considered when designing \acp{IVIS}.

\subsection{Drivers' Self-Regulative Behavior}\label{ch:SelfRegulation}
While interacting with touch-based \acp{IVIS}, drivers divide their visual attention between the primary driving task and the secondary touchscreen interaction. Research shows that drivers actively self-regulate their multitasking behavior to maintain safe driving. They adapt their level of engagement to mitigate the risks associated with the secondary task demands~\citep{rudin-brown.2013a}. According to~\citet{rudin-brown.2013a}, this self-regulative behavior can be intentional or unintentional. The authors further argue that it occurs at three distinct levels derived from Michon's driver task model~\citep{michon.1985}: strategic, tactical, and operational. 

Strategic self-regulation describes driver decisions that are made on a timescale of minutes or more~\citep{rudin-brown.2013a}. These decisions are often constant over a trip. Some drivers, for example, report that they never engage in a secondary task in heavy traffic, in poor weather conditions, or when driving at nighttime~\citep{young.2010}. \citet{oviedo-trespalacios.2019} modeled strategic self-regulation as the decision to pull over to perform a secondary task. In this study, some drivers made the strategic decision to not engage in secondary tasks while driving. 

Tactical self-regulation refers to a driver's decision to engage in a secondary task according to the driving demand. Drivers make tactical decisions in the time frame of seconds~\citep{rudin-brown.2013a} and continuously update them while driving. Many studies investigate drivers' engagement in mobile phone tasks while driving. The results show that drivers are less likely to engage in a visual manual phone task when driving demands are high (high speed, sharp turns, etc.)~\citep{tivesten.2015, hancox.2013, oviedo-trespalacios.2018a, oviedo-trespalacios.2019, ismaeel.2020}. \citet{tivesten.2015} show that drivers use information about the upcoming driving demand to decide whether or not to engage in a secondary task. 
Somewhat contrary results are presented by \citet{horrey.2009}. The authors found that, although drivers were well aware of the demands of specific traffic situations, it had little influence on the decision to engage in the secondary task. This is consistent with findings presented by~\citet{carsten.2017} who show that drivers stopped engaging in easy tasks when the driving demand increased but continued to engage in more demanding secondary tasks. \citet{liang.2015} found that drivers avoided initiating a secondary task before an immediate transition to higher driving demands. However, drivers did not postpone their secondary task engagement when driving demand was already high~\citep{liang.2015}. \citet{carsten.2017} and~\citet{liang.2015} argue that more work is needed to evaluate the factors influencing tactical self-regulation.

Operational self-regulation describes behavioral adaptions made by the driver while actively engaging in a secondary task. This implies that on the strategic and tactical level the driver already decided to engage in a secondary task. Operational self-regulation can be bidirectional. Research shows that drivers adjust their driving behavior in terms of vehicle speed, lane position, or time headway, when they engage in a secondary task~\citep{schneidereit.2017, morgenstern.2020, onate-vega.2020, oviedo-trespalacios.2018,choudhary.2017}. On the other hand, recent findings show that drivers also adjust their secondary task engagement in response to variations in driving demand. \citet{oviedo-trespalacios.2019} found that drivers temporarily stopped the use of mobile phones to cope with varying driving demands~\citep{oviedo-trespalacios.2019}. Similarly, in a test track experiment, \citet{liang.2015} show that drivers adjust their time-sharing behavior according to driving demands~\citep{liang.2015}. In addition, \citet{tivesten.2014} state that drivers allow for more distraction in less demanding situations. In a naturalistic driving study, drivers performed shorter off-road glances during turning when a lead vehicle was present and when they detected oncoming traffic~\citep{tivesten.2014}. \citet{tivesten.2014} further state that drivers prioritize secondary tasks over monitoring the driving environment, especially in low-speed situations. Accordingly, \citet{risteska.2021} show that drivers' off-path glances decrease in situations with higher visual difficulty~\citep{risteska.2021}.

\subsection{The Effect of Driving Automation on Self-Regulation}

Many studies have investigated the effect of partially automated driving (Level 1 and Level 2 according to SAE J3016 ~\citep{SAE.2021}) on drivers' secondary task engagement. As laid out in the following, the results suggest that more automation results in less driver engagement and, thus, a lower capability to correctly assess the current driving situation.

\citet{lin.2019} investigate drivers' self-regulation in Level 2 driving according to the levels of situation awareness as proposed by~\citet{schomig.2013}. On the control level, which corresponds to operational self-regulation, they found that drivers adapt their behavior according to the severity of the hazard. Whereas they pause their engage in case of urgent hazards, they continue to engage with a more frequent task switching behavior) in case of less urgent hazards. In addition, many studies investigated how drivers allocate their visual attention during partially automated driving. Results from the Virginia Connected Corridors Level 2 naturalistic driving study~\citep{dunn.2020} indicate that the use of Level 2 automation (i.e., \ac{ACC}+\ac{LCA}) led to drivers spending less time with their eyes on driving-related tasks. In accordance,~\citet{gaspar.2019} found that with partial automation activated, drivers made longer single off-road glances and had longer maximum total-eyes-off-road times~\citep{gaspar.2019}. This finding is complemented by the results presented by~\citet{yang.2021} who also found that off-road glances were longer in automated driving conditions and additionally investigated the effect of different levels of distraction. They found that off-road glances were longer for highly distracting secondary tasks~\citep{yang.2021}.
\citet{noble.2021} assessed the effect of \ac{ACC}, \ac{LCA}, and \ac{ACC} + \ac{LCA} on drivers' glance behavior and secondary task engagement. They found that during ACC+LCA driving, drivers execute longer and more frequent glances away from the road~\citep{noble.2021}. They, however, did not find significant differences in the mean off-road glance duration nor in the tactical self-regulation when \ac{ACC} + \ac{LCA} was active. Another naturalistic driving study is presented by~\citet{morando.2019} who found a significant decrease in the percentage of time with eyes on the road center when using \ac{ACC} + \ac{LKA}~\citep{morando.2019}. In a subsequent study, the authors investigated drivers' glance behavior during disengagements of Tesla's Autopilot in naturalistic highway driving~\citep{morando.2021}. Whereas they found that all off-road glances tended to be longer with AP compared to manual driving, the difference was particularly big for glances down and toward the center stack. The mean glance duration increased by 0.3 seconds and the proportion of glances longer than 2 seconds increased by 425\,\% in Autopilot conditions compared to manual driving.

\subsection{Research Questions}\label{ch:RQs}
We identify two main research gaps in the current state of the art: (1) Current work is mainly focused on self-regulation when interacting with mobile phones or when engaging in general secondary tasks such as eating, drinking, or talking to a passenger. No work addresses operational and tactical self-regulatory behavior during explicit interactions with \acp{IVIS}. (2) Whereas multiple studies investigate the general effect of partial automation on drivers' self-regulation, there is yet no detailed investigation on the interdependencies between driving automation, vehicle speed, and road curvature.

Considering that modern \acp{IVIS} are increasingly complex and incorporate nearly all the functionality of smartphones and that \ac{ACC} and \ac{LCA} are becoming more capable and accessible, we argue that both aspects need to be examined in more detail. Therefore, we aim to answer the following research questions:

	\begin{itemize}
		\item[\textbf{RQ1:}] To what extent do drivers self-regulate their behavior on the tactical level when engaging in secondary touchscreen tasks depending on driving automation, vehicle speed, and road curvature?
		\item[\textbf{RQ2:}] To what extent do drivers self-regulate their behavior on the operational level when engaging in secondary touchscreen tasks depending on driving automation, vehicle speed, and road curvature?
		\item[\textbf{RQ3:}] Does the effect of driving automation on drivers' operational self-regulation vary in response to different driving situations?
	\end{itemize}

\section{Method}

\begin{figure*}
	\centering
	\includegraphics[width=\linewidth]{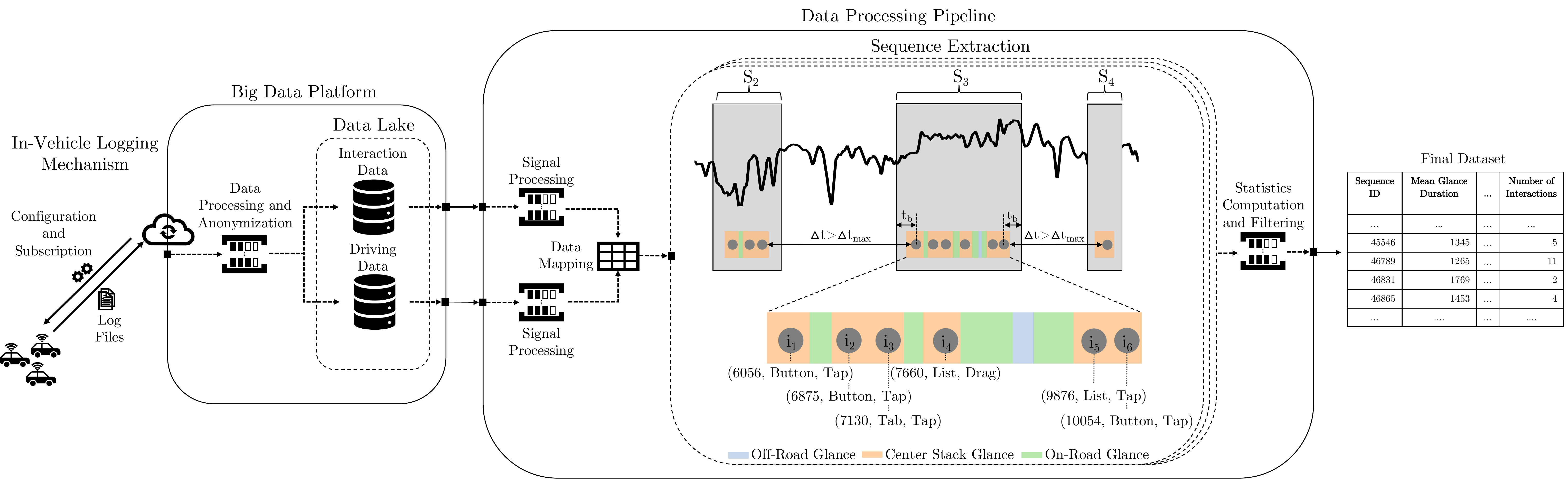} 
	\caption{Schematic overview of the data collection and processing procedure. Adjusted according to \citet{ebel.2021a} and \citet{Ebel.2023}.} 
	\label{fig:DataCollectionAndProcessing} 
\end{figure*}

\subsection{Data Source and Data Collection}
\begin{figure}
	\centering
	\includegraphics[width = 0.8\linewidth]{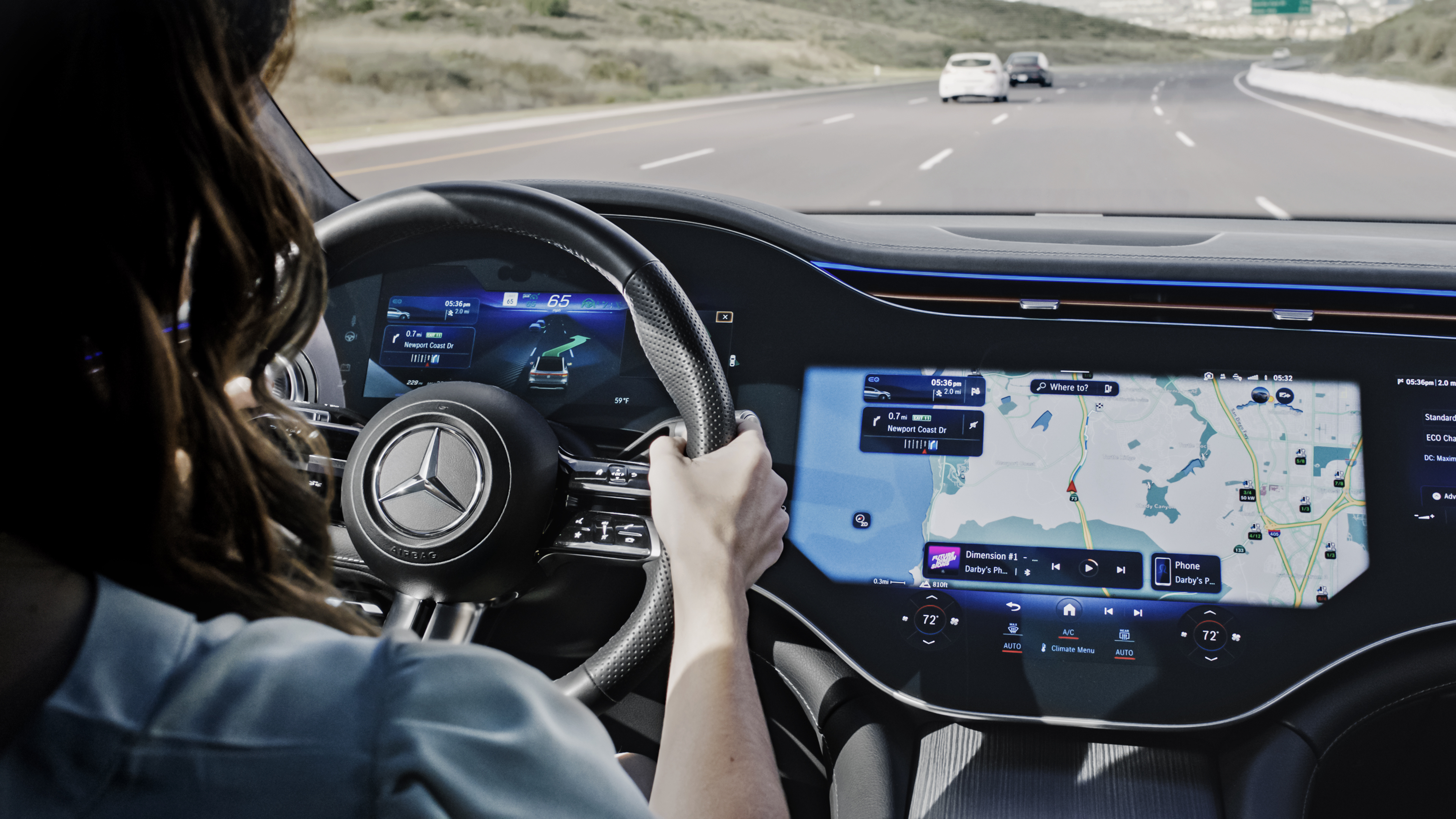} 
	\caption{A center stack touchscreen representative for the touchscreens evaluated in this work~\citep{Mercedes.2023}.} 
	\label{fig:Touchscreen} 
\end{figure}

In this work, we analyze 31,378 interaction sequences extracted from 10,402 individual trips. More than 100 test vehicles and five different car models contributed to the data collection from mid-October 2021 to mid-October 2022. The vehicles are part of the internal test fleet of Mercedes-Benz. Figure \ref{fig:Touchscreen} shows a vehicle with an interior that is representative of the cars in the fleet. They are used for a variety of testing procedures but also for transfer and leisure drives of employees. All vehicles that are equipped with the most recent software architecture, a stereo camera for glance detection, and ACC and LCA technology, contributed to the data collection. ACC automates the longitudinal control and LCA supports the lateral control keeping the car in the center of the lane. Both systems work at speeds between 0\,km/h and 210\,km/h. An additional feature is the so-called active traffic jam assist. If both systems are active and the driver is in a traffic jam on a multi-lane road with separate carriageways, the system can fully control steering and acceleration up to 60\,km/h. However, the driver is still obliged to monitor the driving environment at all times. Thus, it is still a Level 2 driving automation systems according to SAE J3016~\citep{SAE.2021}.

All data used in this work was collected over the air, via the telematics data collection framework of Mercedes-Benz (see \textit{In-Vehicle Logging Mechanism} and \textit{Big Data Platform} in Figure~\ref{fig:DataCollectionAndProcessing}). The \textit{In-Vehicle Logging Mechanism} allows the collection of interaction data via the \ac{HMI} Interface and the collection of driving and camera data via the \ac{CAN}. Once a new configuration file is deployed to a car, the specified datapoints are logged and transferred to the \textit{Big Data Platform}. Here, the data is processed and anonymized. All datapoints that were collected during the same trip are given the same unique identifier. Afterward, interaction and driving data is stored in a data lake. 

In this work, we analyze touchscreen interactions, driving data (vehicle speed, steering wheel angle, and level of driving automation), and eye tracking data. Steering wheel angle and vehicle speed are logged at a frequency of $4\,Hz$. For each user interaction on the center stack touchscreen a data point that consists of a timestamp, the interactive UI element, and the coordinates of the fingers is logged. Based on the name of the UI element, each interaction is mapped to one of the broader categories shown in Table~\ref{tab:UIelements}. The type of gesture that the driver performed is inferred based on the press and release coordinates the detected touch points. To control for touchscreen interactions that are not performed by the driver, we also collect the seat belt signal of the front passenger. This allows us to detect sequences in which a passenger was present and might have interacted with the center stack touchscreen.

The glance data is acquired using a stereo camera located in the instrument cluster behind the steering wheel. The eye tracking is primarily based on the pupil-corneal reflection technique~\citep{merchant.1967}, which is used in the majority of remote eye tracking devices~\citep{hutchinson.1989}. The driver’s field of view is divided into different \acp{AOI} and the system continuously keeps track of the driver’s gaze by mapping it to one of the \acp{AOI}. The true positive rate of the \acp{AOI} describing the center stack touchscreen is above 90 percent. The system used in this research is a production system without the ability to capture raw video data.

\subsection{Data Processing}

After the data is logged, anonymized, and stored, each signal is further processed as outlined below and visualized in Figure~\ref{fig:DataCollectionAndProcessing}. These processing steps were developed in our previous work~\citep{ebel.2022, ebel.2021a}.

\subsubsection{User Interaction Data}
In contrast to controlled experiments, there is no predefined secondary task that the drivers have to perform. We know nothing about the drivers' intentions and do not know, which interactions belong together to perform a certain task. We rather observe drivers' natural behavior in an unbiased setting. We, therefore, extract user interaction sequences based on the assumption that drivers disengaged from the secondary task when they do not interact with the touchscreen for more than $\Delta t_{max} = 10\,\text{s}$ (see Figure~\ref{fig:DataCollectionAndProcessing}). The next interaction is then considered the starting point of a new interaction sequence.

\subsubsection{Eye Tracking Data}
\begin{figure}
	\centering
	\includegraphics[width = 0.8\linewidth]{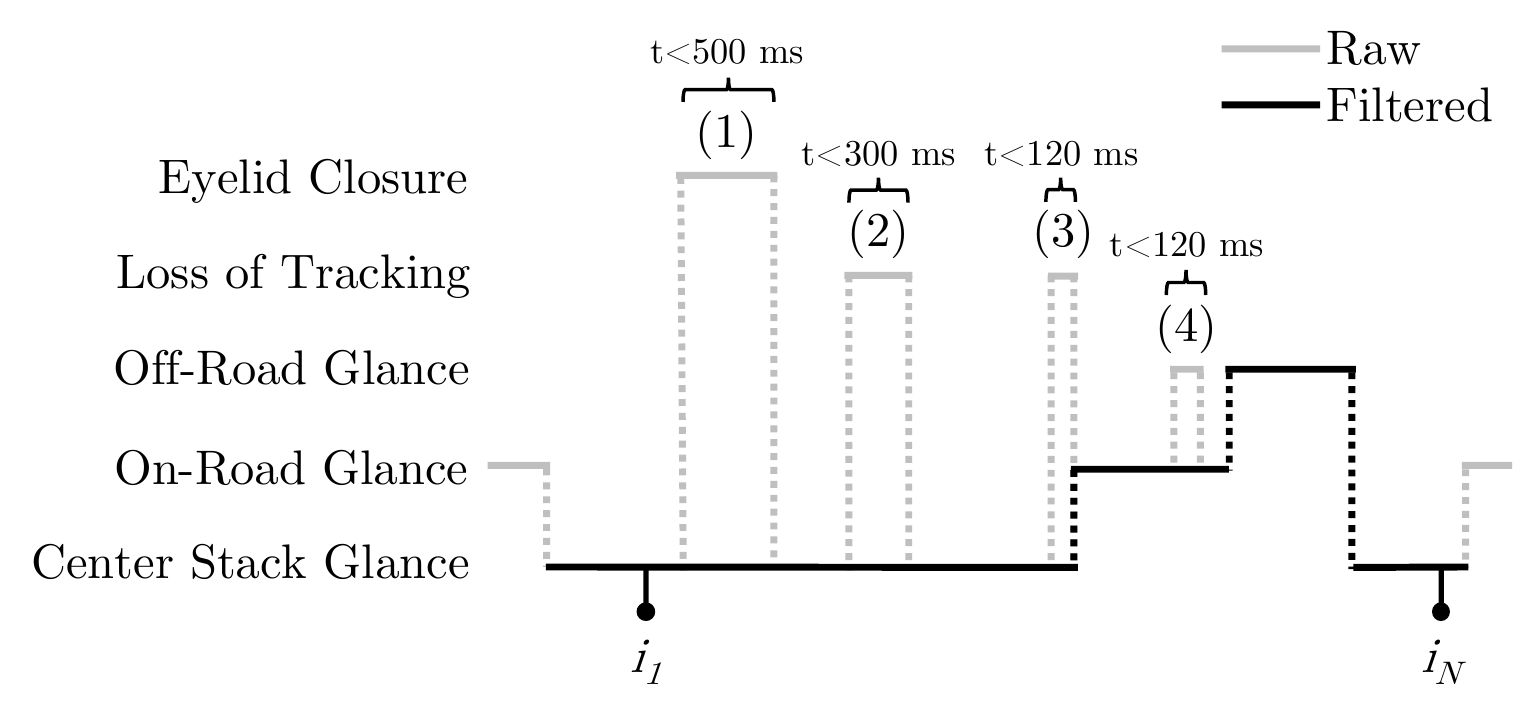} 
	\caption{Glance processing procedure where $i_1$ indicates the first touchscreen interaction of a sequence and $i_N$ indicates the last one. (1) Eyelid closure shorter than 500\,ms, preceding and subsequent AOI are similar (2) Loss of tracking shorter than 300\,ms, preceding and subsequent AOI are similar (3) Loss of tracking shorter than 120\,ms, preceding and subsequent AOI are different (4) Fly-through shorter than 120\,ms, preceding and subsequent AOI are similar} 
	\label{fig:GlanceProcessing} 
\end{figure}

We extract all glances toward the center stack touchscreen between the first $i_1$ and last interaction $i_N$ of each interaction sequence. To improve the quality of the eye tracking data, we apply several filtering steps as depicted in Figure \ref{fig:GlanceProcessing}. The processing is partially adapted from related work~\citep{morando.2019} and follows ISO 15007-1:2020~\citep{ISO15007}. (1) First, we filter all eyelid closures shorter than $500\, ms$ to remove normal blinks and eyelid closures not associated with microsleeps. (2) To handle short periods of tracking loss, we interpolate gaps shorter than $300\,ms$ if the preceding \ac{AOI} is equal to the succeeding one, and (3) gaps shorter than $120\,ms$ if the preceding and succeeding \acp{AOI} are different. $120\,ms$ is the shortest fixation that humans can control~\citep{ISO15007} and shorter fixations are physiologically impossible. Accordingly, to remove fly-throughs, (4) we also interpolate all glances shorter than $120\,ms$. When glances are interpolated, the duration of the filtered glance or tracking loss is added to the duration of the previous AOI if preceding and subsequent AOIs are different (see (3) in~\autoref{fig:GlanceProcessing}). If preceding and subsequent AOI are similar, the surrounding glances are merged as shown in (1) in~\autoref{fig:GlanceProcessing}

\subsubsection{Driving Data}
The driving data consists of vehicle speed, steering wheel angle, and automation level. First, we extract all data that is relevant for a specific interaction sequence. For each sequence, we consider the vehicle speed and steering wheel angle data from two seconds before the first interaction until $t_b = 2\,\text{s}$ after the last interaction (see Figure~\ref{fig:DataCollectionAndProcessing}). This allows to compute more stable aggregate statistics for very short sequences. We discard all sequences for which deviations in the logging frequency or sensor outages were detected.

\subsubsection{Final Filtering and Data Description}\label{ch:filtering}

\begin{figure}
	\centering
	\includegraphics[width = 0.5\linewidth]{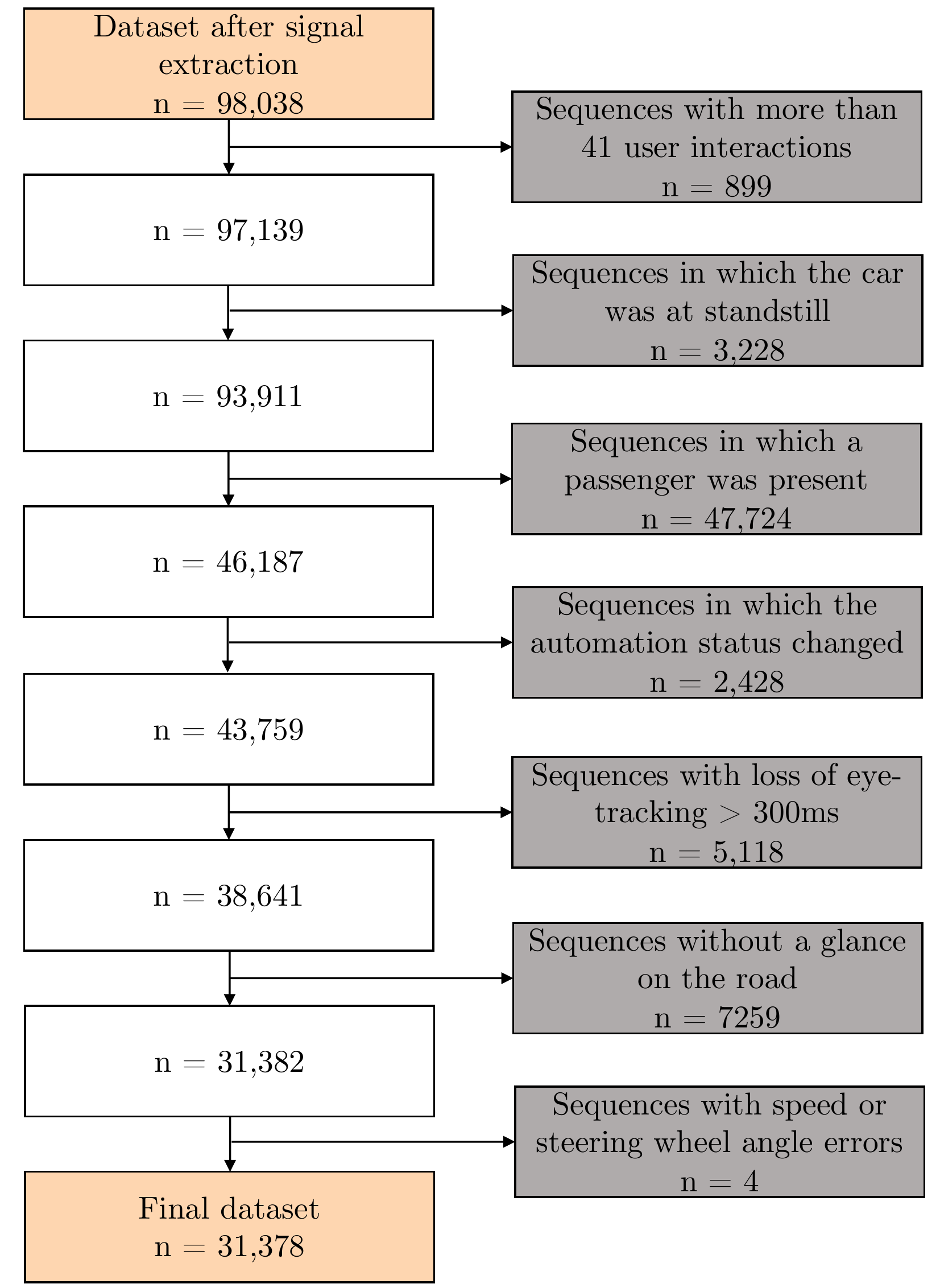} 
	\caption{Data filtering procedure} 
	\label{fig:DataExclusion} 
\end{figure}

After individual signal extraction, the dataset contains 98,038 sequences. To improve data quality and control of confounding factors, we apply strict exclusion criteria as visualized in Figure~\ref{fig:DataExclusion}. We discard all sequences with more than 41 interactions, which corresponds to the $99^{th}$ percentile of the distribution of interactions per sequence. We further discard all sequences in which the car was at standstill. We filter these sequences, because we are only interested in self-regulation while driving. To control for potential distractions or interactions by the front passenger, we delete all sequences in which the front passenger seat belt buckle was latched. We, further, discard all sequences in which the automation level cannot be unambiguously assigned. This includes sequences in which the driver selected another automation level or overwrote the current level by accelerating or braking. The driving automation can also be deactivated due to external factors like a loss of lane marking or bad weather conditions. Furthermore, all sequences with a loss of eye tracking larger than $300\,ms$ are deleted. In contrast to our previous work \citep{ebel.2022}, we also discard all sequences during which the driver did not perform a gaze transition between the center stack touchscreen and the road. As we are interested in drivers' self-regulative behavior, we only consider sequences during which such regulation happened. Lastly, all sequences with errors in the speed or steering wheel angle signal are discarded. The final dataset contains 31,378 sequences of which 18,449 are manual driving, 1,542 are ACC driving, and 11,378 are ACC+LCA driving\footnote{The dataset statistics are given in Appendix~\ref{ch:AppendixSummaryStats}}.

\subsection{Statistical Modeling}
As stated in Section~\ref{ch:RQs}, we investigate how drivers' tactical and operational self-regulation changes in response to different levels of driving automation and driving contexts. In the following, we introduce the dependent and independent variables, and the statistical models we use. We define statistical significance at the level of $\alpha=0.05$.

\subsubsection{Dependent Variables}
We chose following dependent variables to model tactical and operational self-regulation:

\textbf{UI Interactions.} Current approaches are mostly investigating tactical self-regulation by comparing the likelihood of a driver engaging in a specific secondary task given different driving situations. We aim to investigate drivers' tactical self-regulation in greater detail, such that we can draw conclusions about the UI design itself. Therefore, we choose the \textit{number of interactions} (discrete), the \textit{number of touch gestures} (discrete) and the \textit{probability of driver interactions with specific UI elements} (categorical) as dependent variables. The different categories of UI elements and touch gestures are listed in Table~\ref{tab:UIelements}.

\begin{table*}
	\small
	\caption{Overview of the different UI elements and touch gestures used as target variables to model drivers' tactical self-regulation.}
	\label{tab:UIelements}
	\centering
		\begin{tabular}{@{}ll@{}}
		    \toprule
			Category & Description \\
			\toprule
			\textbf{UI Elements} &\\
			Button         & General buttons like push buttons or radio buttons\\
			List           & List containers used, for example, to present destination suggestions\\
			Homebar        & Static homebar on the bottom screens (e.g., music and climate controls) \\
			AppIcon        & Application icons on the home screen, used to start an application\\
			Tab            & Tab bar used to navigate between different views or subtasks\\
			Map            & Map viewer that displays a map and allows for interactions with it \\
			Keyboard       & Virtual keyboard or number pad to enter text \\
            CoverFlow      & Animated widget that, for example, allows flipping through album covers \\
			Slider         & Vertical or horizontal sliders used, for example, when changing the volume \\
			RemoteUI       & Apple Car Play or Android Auto \\
			ControlBar     & Menu controls to show context menus or popups \\
			ClickGuard     & Non-interactive background elements\\
			Other          & UI elements that do not fit any of the above categories \\
			Unknown        & UI elements for which the identifier is not specified \\
			\textbf{Gestures} & \\
			Tap            & A one finger touch on the screen without significant movement\\
			Drag           & A one finger dragging motion\\
			Multitouch     & A multi finger gesture\\
			\bottomrule
		\end{tabular}
\end{table*}

\textbf{Mean Glance Duration.} The \textit{mean glance duration} is a continuous variable. It is computed as the sum of the duration of all glances toward the center stack touchscreen during a sequence divided by the total number of glances per sequence. 

\textbf{Long Glance.} The dichotomous variable \textit{long glance} indicates whether a driver glanced at the center stack touchscreen for more than two seconds. Eyes-off-road glances longer than two seconds are associated with an increased crash risk~\citep{Klauer.2006}. The proportion of such long glances is an important factor in evaluating drivers' operational self-regulation.

\subsubsection{Independent Variables}
The dependent variables are analyzed with respect to the following independent variables:

\textbf{Automation Level.} The automation level is a categorical variable with three distinct levels: \textit{manual, ACC, and ACC+LCA}. According to SAE J3016~\citep{SAE.2021}, these levels correspond to Level 0, 1, and 2 of driving automation. The automation level is constant throughout each sequence.

\textbf{Vehicle Speed.} The vehicle speed is a categorical variable with three levels: $0\,km/h < v \leq 50\,km/h, 50\,km/h < v \leq 100\,km/h\,, v > 100\,km/h$. It is computed as the mean speed across a sequence. 

\textbf{Road Curvature.} The road curvature is a categorical variable with two levels: \textit{straight} or \textit{curved}. An interaction sequence is classified as curved if the maximum absolute steering wheel angle is greater than 50\textdegree or if the absolute mean steering wheel angle is greater than 5\textdegree.

\subsubsection{Models}

To account for the hierarchical data structure and the unbalanced study design we use mixed-effects models. Our data structure is hierarchical because interaction sequences are nested within trips and many trips occur within specific car types. Furthermore, not all combinations of the independent variables are observed in all trips and car types. This results in an unbalanced study design. However, mixed-effects models also referred to as multilevel models~\citep{hox.1998}, are well suited for unbalanced designs and account for grouping hierarchies~\citep{magezi.2015}. Thus they are well suited to test our hypotheses.

We performed all our analyses using R Statistical Software (v4.2.1) \citep{rcoreteam.2022}. We used the \textit{lme4} package (v.1.1.31)~\citep{bates.2015} to build the multilevel models, obtained p-values via the \textit{lmertest} package (v.3.1.3)~\citep{kuznetsova.2017}, and computed the pairwise post-hoc tests using the \textit{emmeans} package (v.1.8.2)~\citep{lenth.2022}. Regression tables were generated using the \textit{stargazer} package (v.5.2.3)~\citep{hlavac.2022}.

\textbf{User Interaction Models.}
To assess tactical self-regulation, we model the driver's decision to engage in a particular task in a particular driving situation. Specifically, we model the probability of drivers interacting with a particular UI element and the number of interactions and gestures drivers perform when interacting with the center stack touchscreen.
To estimate the probability of a driver to engage with one of the UI elements, we fit one logistic mixed-effects model with random intercepts for each type of UI element and type of gestures.
In alignment with our previous work \citep{ebel.2022}, none of the two-way or three-way interactions were significant or proved to significantly improve the predictive performance compared to the additive model. We therefore omit these interaction effects.

To model the number of interactions and gestures that drivers perform during an interaction sequence, we fit two negative binomial mixed-effects models with random intercepts. We use negative binomial models because the number of interactions is a discrete count value. We could have also used Poisson models but our tests have shown that they suffered from overdispersion.

For all user interaction models we include \textit{automation level}, \textit{vehicle speed}, and \textit{road curvature} as fixed effects. Furthermore, we include the trip during which the sequence was recorded and the car type as random effect. 

\textbf{Glance Behavior Models.}
To estimate the \textit{mean glance duration}, we fit four linear mixed-effects models with random intercepts. An exploratory data analysis showed that the distribution of the mean glance duration is heavily right-skewed. To satisfy the model assumption of normally distributed residuals we, therefore, apply a log transformation.
In Model 1 we estimate the effect of driving automation on the mean glance duration across all driving situations by only selecting the automation level as a fixed effect. To account for the hierarchical structure of our data we include the trip during which an interaction sequence was recorded and the car type as random effects for both models. In Model 2 we add the vehicle speed and road curvature as additional fixed effects and allow for interaction effects. Similar to Model 1, the trip and car type are included as random effects. To estimate drivers' long glance probability, we fit two logistic mixed effect models with random intercepts. In Model 3 we select the automation level as a fixed effect and in Model 4 we add the vehicle speed and road curvature as fixed effects and model all interactions between the independent variables. The trip and car type information are, again, entered as random effects.

Visual inspection of residual plots and Q-Q plots of the final models did not reveal any obvious deviations from homoscedasticity or normality. We use Satterthwaite's degrees of freedom approximation to obtain p-values and evaluate significances~\citep{luke.2017}. For the post-hoc pairwise comparisons we use Tukey's multiple comparison method~\citep{tukey.1949}.

\section{Results}
In the following, we present the results obtained by fitting the above-introduced models to the 31,378 interaction sequences. By doing so we can model tactical and operational self-regulation. The analysis of the model coefficients and post-hoc tests allow us to quantify how drivers adapt their multitasking behavior according to changes in speed, road curvature, and driving automation.

\subsection{Tactical Self-Regulation}
\begin{figure*}
	\centering
	\includegraphics[width=\linewidth]{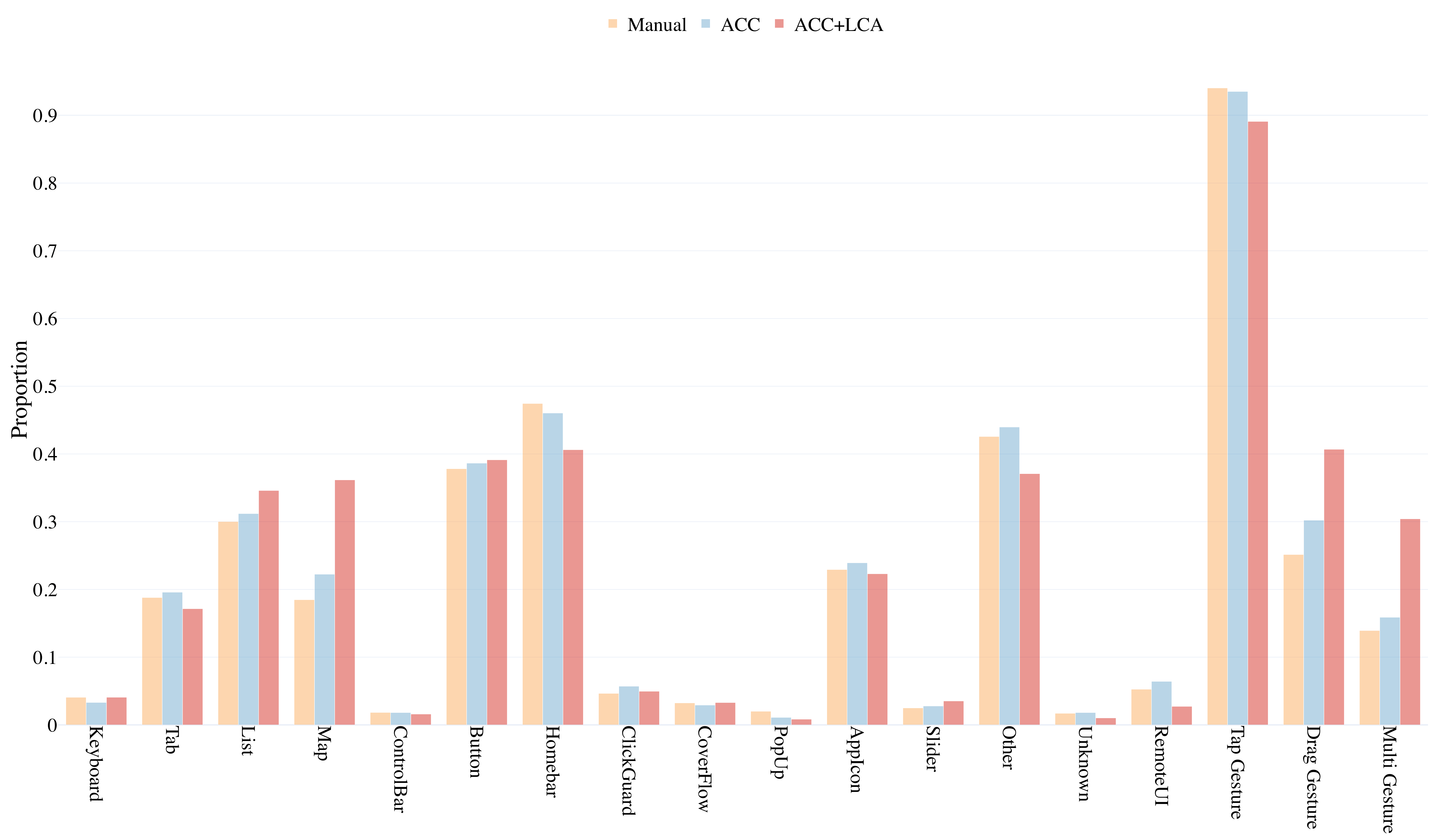} 
	\caption{Proportion of sequences in which the driver interacted with a respective UI element (a) and performed a specific gesture (b).} 
	\label{fig:ElementsAndGesture} 
\end{figure*}

\subsubsection{Number of Touch Interactions and Touch Gestures}

\begin{sidewaystable}[!htbp]
    \centering 
  \caption{Negative binomial mixed-effects models describing the number of touchscreen interactions, tap gestures, drag gestures, and multitouch gestures during an interaction sequence. For each model, the intercept and the coefficients describe the effect of the independent variables. They are shown along with the estimated standard error. The coefficients and standard errors of the negative binomial mixed-effects model are given on a logarithmic scale.}
\begin{tabular}{@{\extracolsep{5pt}}lrrrr} 
\\[-1.8ex]\hline 
\hline \\[-1.8ex] 
 & \multicolumn{4}{c}{\textit{Dependent variable:}} \\ 
\cline{2-5} 
\\[-1.8ex] & Num. Interactions & Num. Tap Gestures & Num. Drag Gestures & Num. Multitouch Gestures \\ 
\hline \\[-1.8ex] 
Intercept & 1.74$^{***}$ (0.01) & 1.52$^{***}$ (0.02) & $-$1.12$^{***}$ (0.03) & $-$2.77$^{***}$ (0.09) \\
\multicolumn{5}{@{} l}{\textbf{Automation Level}}\\
  \hspace{2mm} Manual$^\dagger$ &  & & &  \\ 
  \hspace{2mm} ACC & 0.11$^{***}$ (0.02) & 0.07$^{**}$ (0.03) & 0.29$^{***}$ (0.06) & 0.28$^{**}$ (0.09) \\ 
  \hspace{2mm} ACC+LCA & 0.16$^{***}$ (0.01) & 0.07$^{***}$ (0.01) & 0.55$^{***}$ (0.03) & 0.47$^{***}$ (0.05) \\ 
  \multicolumn{5}{@{} l}{\textbf{Vehicle Speed}}\\
  \hspace{2mm} 0-50$^\dagger$ &  &  &  &  \\ 
  \hspace{2mm} 50-100 & 0.04$^{***}$ (0.01) & 0.03$^{*}$ (0.01) & 0.12$^{***}$ (0.03) & 0.03 (0.04) \\ 
  \hspace{2mm} 100+ & 0.01 (0.01) & $-$0.01 (0.01) & 0.15$^{***}$ (0.04) & 0.04 (0.05) \\
  \multicolumn{5}{@{} l}{\textbf{Road Curvature}}\\
  \hspace{2mm} straight$^\dagger$ &  & &  &  \\ 
  \hspace{2mm} curved & $-$0.17$^{***}$ (0.01) & $-$0.13$^{***}$ (0.01) & $-$0.42$^{***}$ (0.04) & $-$0.33$^{***}$ (0.05) \\ 
 \hline \\[-1.8ex] 
Akaike Inf. Crit. & 173,027.10 & 163,962.60 & 69,556.59 & 53,842.50 \\ 
Bayesian Inf. Crit. & 173,102.30 & 164,037.80 & 69,631.77 & 53,917.68 \\ 
\hline 
\hline \\[-1.8ex] 
\textit{Note:}  & \multicolumn{4}{r}{$^\dagger$ indicates the reference group, $^{*}$p$<$0.05; $^{**}$p$<$0.01; $^{***}$p$<$0.001} \\ 
\end{tabular} 
\label{tab:InteractionsAndGestureModel}
\end{sidewaystable}

Table~\ref{tab:InteractionsAndGestureModel} shows the parameters of the user interaction models. We modeled the number of touch interactions, tap, drag, and multitouch gestures per sequence. The results suggest that driving automation, vehicle speed, and road curvature affect the number of touchscreen interactions and gestures that drivers perform when engaging with the center stack touchscreen. The influence of the independent variables is generally similar but differs significantly in magnitude comparing \textit{Tap} gestures to \textit{Drag} and \textit{Multitouch} gestures.

The $\beta$ coefficients of the negative binomial model are given on a logarithmic scale. They can be interpreted as follows: Keeping everything else constant, an increase of one level in the predictor variable results in a $e^{\beta}$ increase of the dependent variable.
Thus, drivers perform $e^{0.11} \approx 1.12$ as many interactions during ACC driving and $e^{0.16} \approx 1.17$ as many interactions during ACC+LCA driving compared to manual driving. This corresponds to an increase of 12\,\% and 17\,\% respectively. Considering the different gestures that add up to the number of interactions, the modeling results suggest that, during automated driving, drivers in particular perform more drag or touch gestures compared to regular tap gestures. For example, during ACC+LCA driving the number of \textit{Tap} gestures per sequence increases by 7\,\% whereas the number of \textit{Drag} and \textit{Multitouch} gestures increases by 73\,\% and 60\,\% respectively. 

Road curvature also significantly affects the number of interactions and gestures that drivers perform on the center stack touchscreen. During curved driving, drivers perform $e^{-0.17} \approx 0.84$ as many interactions compared to straight driving. Wheres they only perform 12\,\% less  \textit{Tap} gestures, the number of \textit{Drag} and \textit{Multitouch} gestures reduces by 34\,\% and 28\,\% respectively.

The effect of the vehicle speed on the number of interactions and gestures is in general smaller compared to the effect of driving automation and road curvature. The results indicate that drivers do not, or only slightly, adapt their tap and multitouch behavior in response to changes in vehicle speed. However, the number of \textit{Drag} gestures that drivers perform is significantly higher when driving at speeds above 50\,km/h compared to driving at speeds of 50\,km/h and below.

\subsubsection{Type of UI Elements}
Table~\ref{tab:InteractionResults} shows the parameters of the user interaction models for all UI elements that occur in more than 10\,\% of all sequences\footnote{The results of the other models are provided in the Appendix \ref{ch:AppendixModels}}. The models were fit to predict the probability that a driver interacts with a specific UI element given the automation level, vehicle speed and road curvature. The results suggest that drivers adapt their interaction behavior with the center stack touchscreen based on automation status, vehicle speed, and road curvature. However, these effects do significantly differ for different types of UI elements.

\begin{sidewaystable}
\centering 
  \caption{Generalized linear mixed-effects models describing the probability of the driver interacting with Tab, List, Button, Homebar, or AppIcon UI elements during an interaction sequence. For each model, the intercept and the coefficients describe the effect of the independent variables. They are shown along with the estimated standard error. Coefficients and standard errors correspond to log odds ratios.} 
  \label{tab:InteractionResults} 
\begin{tabular}{@{\extracolsep{0pt}}lrrrrrr} 
\\[-1.8ex]\hline 
\hline \\[-1.8ex] 
 & \multicolumn{6}{c}{\textit{Dependent variable:}} \\ 
\cline{2-7} 
\\[-1.8ex] & Tab & List & Map & Button & Homebar & AppIcon \\ 
\hline \\[-1.8ex] 
 Intercept & $-$1.60$^{***}$ (0.05) & $-$0.98$^{***}$ (0.05) & $-$2.40$^{***}$ (0.09) & $-$0.52$^{***}$ (0.04) & $-$0.04 (0.05) & $-$1.37$^{***}$ (0.05) \\
 \multicolumn{7}{@{} l}{\textbf{Automation Level}}\\
  \hspace{2mm} Manual$^\dagger$ &  &  &  &  &  &  \\ 
  \hspace{2mm} ACC & 0.12 (0.08) & 0.06 (0.07) & 0.18 (0.10) & 0.05 (0.06) & $-$0.01 (0.07) & 0.04 (0.08) \\ 
  \hspace{2mm} ACC+LCA & $-$0.07 (0.04) & 0.25$^{***}$ (0.04) & 0.48$^{***}$ (0.05) & 0.15$^{***}$ (0.03) & $-$0.17$^{***}$ (0.04) & $-$4 (0.04) \\ 
   \multicolumn{7}{@{} l}{\textbf{Vehicle Speed}}\\
  \hspace{2mm} 0-50$^\dagger$ &  &  &  &  &  &  \\ 
  \hspace{2mm} 50-100 & $-$0.01 (0.04) & 0.08$^{*}$ (0.03) & 0.07 (0.05) & 0.02 (0.03) & $-$0.01 (0.03) & 0.06 (0.04) \\ 
  \hspace{2mm} 100+ & $-$0.18$^{***}$ (0.05) & 0.09$^{*}$ (0.04) & 0.13$^{*}$ (0.06) & $-$0.04 (0.04) & $-$0.08 (0.04) & 0.02 (0.04) \\ 
   \multicolumn{7}{@{} l}{\textbf{Road Curvature}}\\
   \hspace{2mm} straight$^\dagger$ &  &  &  &  &  &  \\ 
  \hspace{2mm} curved & $-$0.10$^{*}$ (0.04) & $-$0.13$^{***}$ (0.04) & $-$0.26$^{***}$ (0.05) & $-$0.02 (0.03) & $-$0.03 (0.04) & $-$0.09$^{*}$ (0.04) \\ 
 \hline \\[-1.8ex] 
Akaike Inf. Crit. & 29,284.14 & 38,230.86 & 29,078.23 & 41,130.21 & 41,482.00 & 32,765.17 \\ 
Bayesian Inf. Crit. & 29,350.98 & 38,297.69 & 29,145.06 & 41,197.04 & 41,548.83 & 32,832.00 \\ 
\hline 
\hline \\[-1.8ex] 
\textit{Note:}  & \multicolumn{6}{r}{$^\dagger$ indicates the reference group, $^{*}$p$<$0.05; $^{**}$p$<$0.01; $^{***}$p$<$0.001} \\ 
\end{tabular}
\end{sidewaystable}

The $\beta$ coefficients for the independent variables given in Table~\ref{tab:InteractionResults} represent log-odds ratios. This means that, keeping everything else constant, a change in the predictor by one level results in a $e^{\beta}$ increase or decrease in the odds that the driver interacts with the respective UI element. Considering the \textit{Map} model the coefficients can be interpreted as follows: During ACC+LCA  driving the odds that a driver performs a map interaction are $e^{0.48} \approx 1.62$ times as high as the odds of performing the same interaction in manual driving. On the other hand, when driving in curved conditions, the odds that the driver interacts with the map are $e^{-0.26} \approx 0.77$ the odds of performing a map interaction in straight driving conditions.

Whereas the effect of ACC isn't significant for any of the models, the effect of ACC+LCA is significant for all Models except of the \textit{Tab} and \textit{AppIcon} models. While drivers are more likely to interact with \textit{List}, \textit{Map}, and \textit{Button} elements, they are less likely to interact with the \textit{Homebar}. The odds to interact with the homebar are $e^{-0.17} \approx 0.84$ the odds compared to manual driving. These effects are also shown in Figure~\ref{fig:ElementsAndGesture}.

Concerning the effect of vehicle speed, the effect of 50--100 is only significant for \textit{List} interactions, suggesting that drivers perform more list interactions when driving between 50\,km/h and 100\,km/h compared to driving at speeds equal to or below 50\,km/h. The effect of 100+ is, however significant for \textit{Tab}, \textit{List}, and \textit{Map}. Whereas the odds of drivers interacting with Tab elements are $e^{-0.18} \approx 0.84$ times the odds of performing the same interactions at speeds between 0\,km/h and 50\,km/h. In contrast, for \textit{List} and \textit{Map} interactions the odds are 1.09 and 1.14 times higher.

The effect of road curvature is significant in all models but the \textit{Homebar} and \textit{Button} models. The coefficients suggest that during curved driving, drivers are in general less likely to interact with the center stack touchscreen. The odds for a driver to interact with these elements in curved driving conditions are between $0.77$ and $0.91$ the odds compared to straight driving.

\begin{figure*}%
    \centering
    \begin{subfigure}{0.32\linewidth}
    \includegraphics[width=\linewidth]{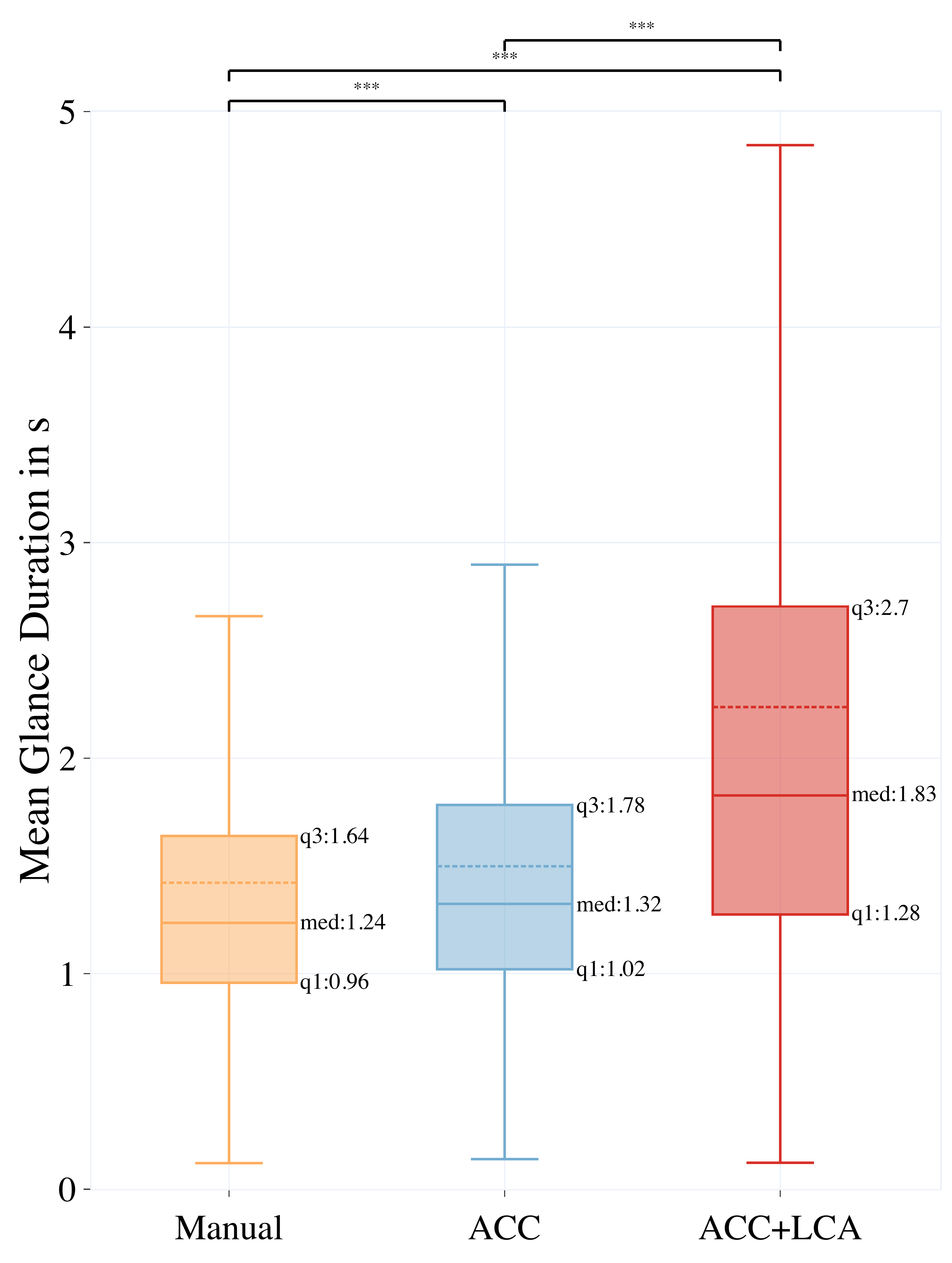}
    \caption{Driving Automation}
    \label{fig:Boxplot_Mean_Glance_Automation}
    \end{subfigure}
    \begin{subfigure}{0.32\linewidth}
    \includegraphics[width = \linewidth]{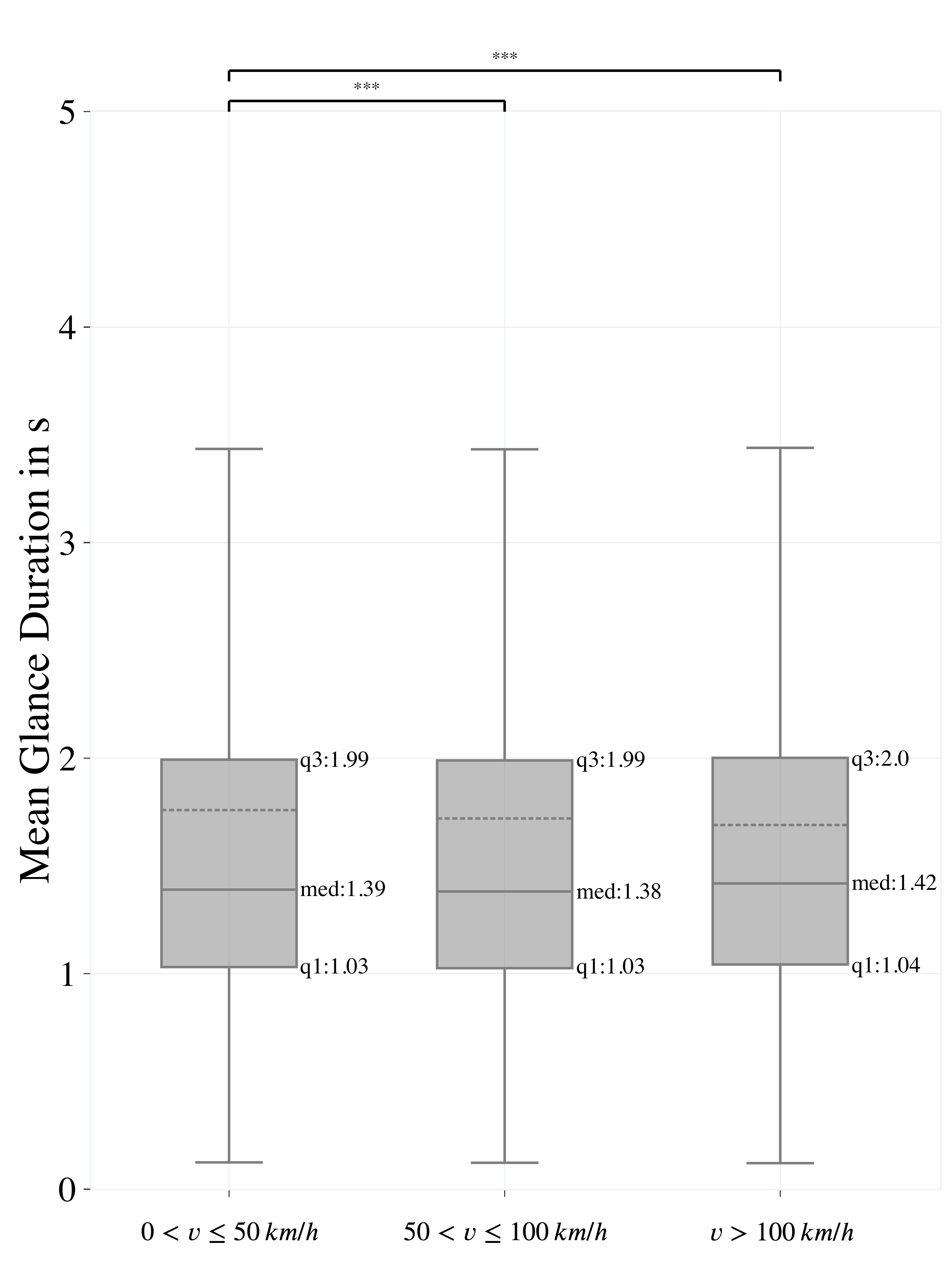}
    \caption{Vehicle Speed}
    \label{fig:Boxplot_Mean_Glance_Speed}
    \end{subfigure}
    \begin{subfigure}{0.32\linewidth}
    \includegraphics[width = \linewidth]{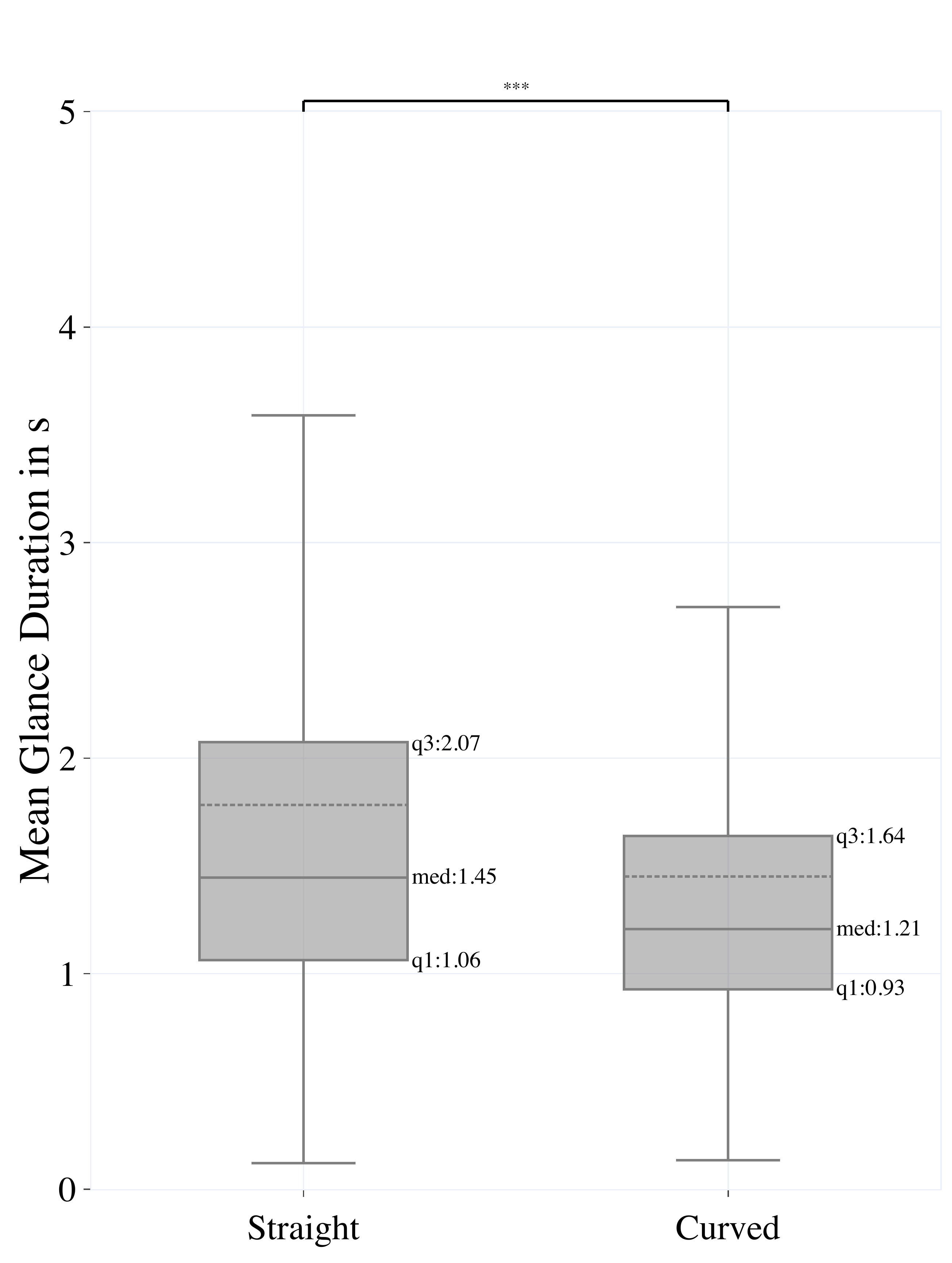}
    \caption{Road Curvature}
    \label{fig:Boxplot_Mean_Glance_Curvature}
    \end{subfigure}
    \caption{Boxplots of the mean glance duration toward the center stack touchscreen grouped according to the driving automation, vehicle speed, and road curvature. Statistically significant differences according to Tukey's pairwise post-hoc test are indicated as: $^{*}$p$<$0.05; $^{**}$p$<$0.01; $^{***}$p$<$0.001}
    \label{fig:Boxplot_Mean_Glance}
\end{figure*}

Across all models, our results suggest that the effect of ACC+LCA driving on tactical self-regulation is larger than the effect of vehicle speed or road curvature. Whereas the tendencies for ACC driving are similar, the effect proves to be not significant ($p>0.05$ for all models). Furthermore the effect of ACC+LCA driving is largest for list and map interactions and small or even negative for the other UI elements.

\subsection{Operational Self-Regulation}
Operational self-regulation is evaluated by identifying how drivers adapt their glance behavior. We measure glance behavior in terms of mean glance duration and long glance probability. The results of our (generalized) linear mixed-effects models (see Table~\ref{tab:GlanceResults}) suggest that drivers adapt their glance behavior while interacting with the center stack touchscreen based on automation status, vehicle speed, and road curvature.

\begin{table*} \centering
\footnotesize
  \caption{Mixed-effects models for mean glance duration and long glance probability toward the center stack touchscreen. The coefficients and standard errors of the mean glance duration models are given on a logarithmic scale. The coefficients and standard errors for the long glance model represent log odds. All coefficients are shown along with the estimated standard error.} 
  \label{tab:GlanceResults} 
\begin{tabular}{@{\extracolsep{5pt}}lrrrr} 
\\[-1.8ex]\hline 
\hline \\[-1.8ex] 
 & \multicolumn{4}{c}{\textit{Dependent variable:}} \\ 
\cline{2-5} 
\\[-1.8ex] & \multicolumn{2}{c}{Mean Glance Duration} & \multicolumn{2}{c}{Long Glance} \\ 
\\[-1.8ex] & \multicolumn{2}{c}{\textit{linear}} & \multicolumn{2}{c}{\textit{generalized linear}} \\ 
 & \multicolumn{2}{c}{\textit{mixed-effects}} & \multicolumn{2}{c}{\textit{mixed-effects}} \\ 
\\[-1.8ex] & Model 1 & Model 2 & Model 3 & Model 4\\ 
\hline \\[-1.8ex] 
 Constant & 7.15$^{***}$ (0.01) & 7.25$^{***}$ (0.01) & $-$0.25$^{***}$ (0.05) & 0.21$^{***}$ (0.06) \\ 
  ACC & 0.10$^{***}$ (0.01) & 0.03 (0.04) & 0.44$^{***}$ (0.07) & 0.11 (0.19) \\ 
  ACC+LCA & 0.31$^{***}$ (0.01) & 0.39$^{***}$ (0.02) & 1.29$^{***}$ (0.04) & 1.29$^{***}$ (0.09) \\ 
  50-100 &  & $-$0.11$^{***}$ (0.01) &  & $-$0.48$^{***}$ (0.05) \\ 
  100+ &  & $-$0.17$^{***}$ (0.01) &  & $-$0.66$^{***}$ (0.06) \\ 
  curved &  & $-$0.09$^{***}$ (0.01) &  & $-$0.55$^{***}$ (0.06) \\ 
  ACC:50-100 &  & 0.12$^{**}$ (0.04) &  & 0.39 (0.22) \\ 
  ACC+LCA:50-100 &  & $-$0.04$^{*}$ (0.02) &  & 0.14 (0.10) \\ 
  ACC:100+ &  & 0.15$^{***}$ (0.04) &  & 0.68$^{**}$ (0.21) \\ 
  ACC+LCA:100+ &  & $-$0.08$^{***}$ (0.02) &  & 0.17 (0.11) \\ 
  ACC:curved &  & $-$0.03 (0.06) &  & 0.34 (0.33) \\ 
  ACC+LCA:curved &  & $-$0.15$^{***}$ (0.04) &  & $-$0.54$^{**}$ (0.19) \\ 
  50-100:curved &  & $-$0.03 (0.02) &  & $-$0.02 (0.09) \\ 
  100+:curved &  & 0.01 (0.03) &  & $-$0.35$^{*}$ (0.18) \\ 
  ACC:50-100:curved &  & $-$0.07 (0.08) &  & $-$0.96$^{*}$ (0.43) \\ 
  ACC+LCA:50-100:curved &  & 0.07 (0.04) &  & 0.05 (0.23) \\ 
  ACC:100+:curved &  & $-$0.11 (0.10) &  & $-$0.74 (0.56) \\ 
  ACC+LCA:100+:curved &  & 0.10 (0.06) &  & 0.87$^{**}$ (0.30) \\ 
 \hline \\[-1.8ex] 
Akaike Inf. Crit. & 42,903.57 & 42,246.95 & 38,850.68 & 38,392.18 \\ 
Bayesian Inf. Crit. & 42,953.69 & 42,422.38 & 38,892.45 & 38,559.26 \\ 
\hline 
\hline \\[-1.8ex] 
\textit{Note:}  & \multicolumn{4}{r}{$^{*}$p$<$0.05; $^{**}$p$<$0.01; $^{***}$p$<$0.001} \\ 
\end{tabular} 
\end{table*}

\subsubsection{Mean Glance Duration}\label{ch:resultsAverageGlance} 

The results of Model 1 as shown in Table~\ref{tab:GlanceResults} suggest that the effect of ACC and ACC+LCA on drivers' mean glance duration toward the center stack touchscreen is significant ($p<0.001$) compared to manual driving. As the mean glance duration is measured on a logarithmic scale, the exponent of models’ coefficients can be interpreted roughly as percent changes. When ACC is active, drivers' mean glance duration increases by $e^{0.10} \approx 1.11 = 11\,\%$. When ACC and LCA are both active, drivers' mean glance duration increases by $36\,\%$ compared to manual driving. Post-hoc testing using Tukey's pairwise post-hoc tests reveals that the difference between ACC and ACC+LCA is also significant. The effects are shown in Figure~\ref{fig:Boxplot_Mean_Glance_Automation}. Figure \ref{fig:Boxplot_Mean_Glance} also shows the mean glance duration for different speed ranges (Figure~\ref{fig:Boxplot_Mean_Glance_Speed}) and road curvature (Figure~\ref{fig:Boxplot_Mean_Glance_Curvature}). According to the modeling results (see Model 5 Table~\ref{tab:SpeedCurvatureOnly} in the Appendix), drivers' mean glance duration decreases by 6\,\% when driving between 50\,km/h and 100\,km/h and by 8\,\% when driving faster than 100\,km/h compared to driving between 0\,km/h and 50\,km/h. It needs to be noted, that whereas these differences are statistically significant ($p<0.001$) they are not observable in Figure~\ref{fig:Boxplot_Mean_Glance_Speed}. This is because most of the correlation in the data is explained by the combination of fixed and random effects (trip and car type) rather than by the fixed effect (vehicle speed) alone. This means that the effect of the vehicle speed is only significant when taking into account trip and car type information. However, Figure~\ref{fig:Boxplot_Mean_Glance_Speed} only shows the mean glance duration according to the vehicle speed. Our results further show that most of the variance in the data is explained by variations in the trip identifier. Considering that vehicle speeds of 0-50\,km/h occur in urban driving but also in very controlled scenarios in a traffic jam on the highway, the trip identifier might be a proxy for different kinds of trips. This also shows that vehicle speed alone might not be the best indicator for changes in driving demand.

\begin{figure*}
	\centering
	\includegraphics[width=\linewidth]{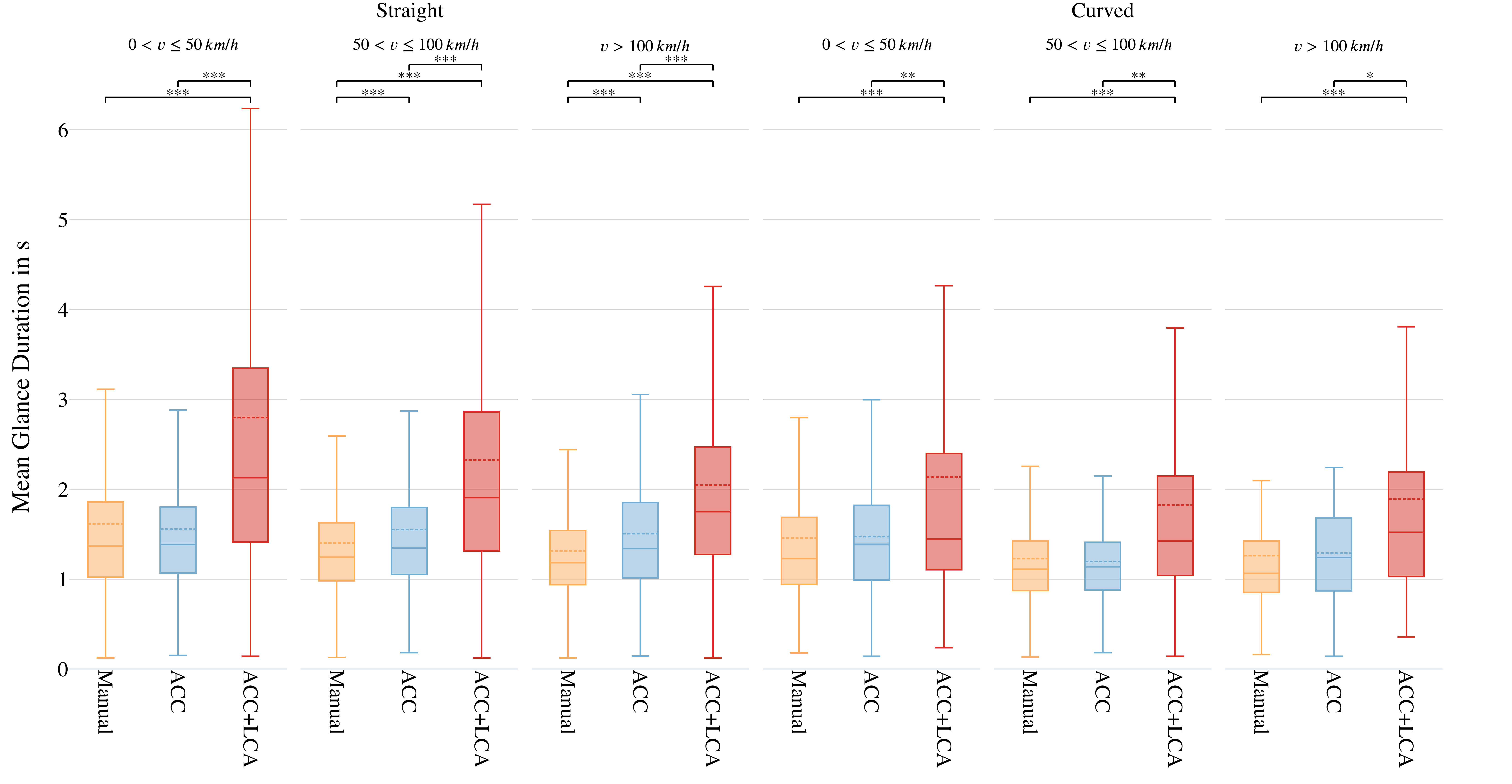} 
	\caption{Boxplots of the mean glance duration toward the center stack touchscreen grouped according to road curvature (left and right half), vehicle speed (combination of three boxplots each), and driving automation (by color). Statistically significant differences according to Tukey's pairwise post-hoc test are indicated as: $^{*}$p$<$0.05; $^{**}$p$<$0.01; $^{***}$p$<$0.001} 
	\label{fig:boxplot_avg_glance_all} 
\end{figure*}

In addition to Model 1, Model 2 adds vehicle speed, road curvature, and the accompanying interactions as fixed effects. In this model, the combination of manual and straight driving, at speeds between $0-50\,km/h$ serves as a reference and all coefficients displayed in Table~\ref{tab:GlanceResults} need to be interpreted accordingly. Apart from the significant main effects for ACC+LCA, 50--100, 100+, and curved, the interactions between both levels of driving automation and vehicle speed and the interaction between ACC+LCA and curved are significant. Whereas the interaction effects of ACC and vehicle speed while driving straight are positive, they are slightly negative for ACC+LCA and vehicle speed. This means that the effect of ACC+LCA decreases slightly for higher speeds during straight driving whereas the effect of ACC increases with the speed for straight sequences. This can also be observed in Figure~\ref{fig:boxplot_avg_glance_all}.

Furthermore, we are interested in whether the effect of ACC and ACC+LCA driving on drivers' self-regulation differs depending on the driving situations. We, therefore, perform pairwise post-hoc comparisons as shown in Figure~\ref{fig:boxplot_avg_glance_all}. We adjust p-values based on Tukey's method for comparing a family of three estimates.

Drivers' mean glance duration is significantly higher during ACC+LCA driving compared to manual driving and ACC driving across all driving situations. During straight driving the mean glance duration during ACC+LCA driving compared to manual driving increases by 47\,\% (0-50\,km/h), 42\,\% (50--100\,km/h), and 36\,\% (100+\,km/h). A similar but slightly smaller effect can be observed during curved driving. Here the mean glance duration increases by 27\,\% (0-50\,km/h), 30\,\% (50--100\,km/h), and 29\,\% (100+\,km/h).

The effect of ACC driving compared to manual driving is only significant for straight driving sequences at speeds between 50\,km/h to 100\,km/h and at speeds above 100\,km/h. For these two conditions drivers' mean glance duration increases by 15\,\% and 19\,\% respectively. During curved driving no significant effect can be observed for ACC driving.

\subsubsection{Long Glance Probability}
The results of Model 3, as presented in Table~\ref{tab:GlanceResults}, suggest that the level of driving automation significantly affects the probability that a driver performs a long glance during an interaction sequence. Both, ACC and ACC+LCA lead to an increase in the long glance probability.

\begin{figure*}
	\centering
	\includegraphics[width=\linewidth]{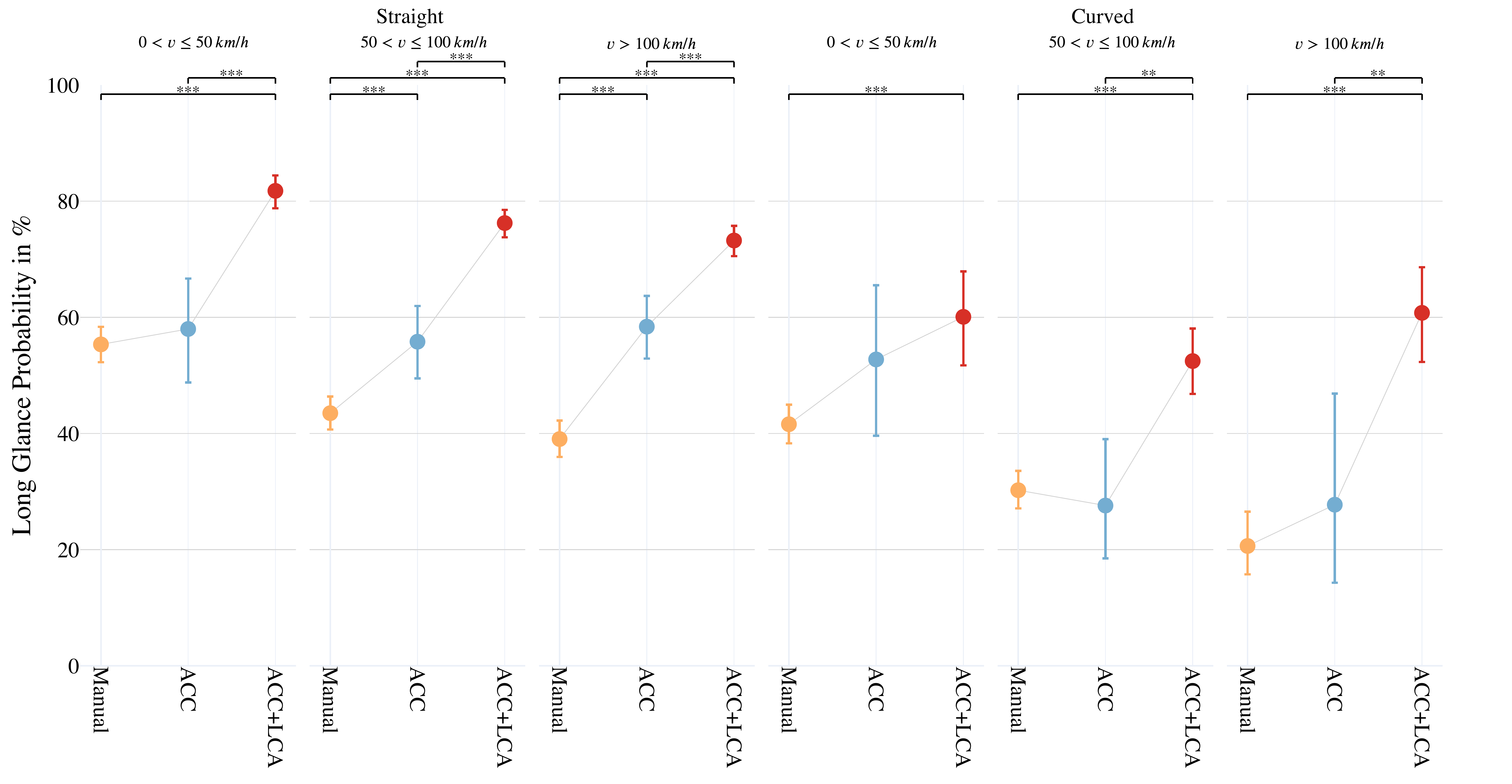} 
	\caption{Marginplot of the predicted long glance probabilities and accompanying confidence intervals. The plots are grouped according to road curvature (left and right half), vehicle speed (combination of three boxplots each), and driving automation (by color). Significant results according to Tukey's pairwise post-hoc test are indicated as: $^{*}$p$<$0.05; $^{**}$p$<$0.01; $^{***}$p$<$0.001} 
	\label{fig:LongGlancePredictions} 
\end{figure*}

The odds that a driver performs a long glance toward the center stack touchscreen are $e^{0.44} \approx 1.6$ (ACC) and $e^{1.29} \approx 3.6$ (ACC+LCA) times higher compared to manual driving. Post-hoc pairwise comparisons also reveal a significant difference between ACC and ACC+LCA with the odds being 2.4 times higher ($p<1$) in the ACC+LCA condition. The results of Model 4 show significant effects of vehicle speed, road curvature, and various interactions. Comparing the main effects, we observe that compared to the reference, the effect of ACC+LCA is roughly twice as high as the effects of 50--100, 100+, or curved. Furthermore, the effect of ACC driving alone is not significant but various of its interaction effects are.
The model predictions and confidence intervals are visualized in Fig~\ref{fig:LongGlancePredictions}. Post-hoc tests comparing the different levels of driving automation for the different combinations of speed and road curvatures were performed using Tukey's multiple comparison method.

For interactions during straight driving at 0-50\,km/h we observe a significant increase ($p<0.001$) in the long glance probability during ACC+LCA driving compared to manual driving and ACC driving. The difference between manual driving and ACC driving is not significant. However, when driving at speeds between 50--100\,km/h and speeds above 100\,km/h, the long glance probability is significantly higher in ACC and ACC+LCA driving compared to manual driving. Similar to the 0-50\,km/h condition the long glance probability during ACC+LCA driving is also significantly higher than during ACC driving.

Considering curved driving conditions, there is no significant difference in the long glance probability between manual and ACC driving across all speed conditions. However, 
for curved driving at speeds of  0-50\,km/h we observe a significant increase in the long glance probability during ACC+LCA driving compared to manual driving (see Figure~\ref{fig:LongGlancePredictions}). For speeds of 50--100\,km/h and speeds above 100\,km/h, the increase in the long glance probability during ACC+LCA is significant compared to both, manual driving and ACC driving. For curved driving, no significant difference can be observed between manual driving and ACC driving. We can also observe that the confidence intervals for all ACC conditions are more widespread compared to the manual driving and ACC+LCA driving conditions.

Also shown in Figure~\ref{fig:LongGlancePredictions}, is the tendency that the long glance probability decreases with an increase in vehicle speed. This is in line with the model coefficients reported for Model 4 in Table~\ref{tab:GlanceResults}. The same holds true for curved driving. Post-hoc pairwise comparisons show that during curved driving drivers' long glance probability decreases significantly across all conditions except ACC driving at speeds 0-50\,km/h ($p=0.5077$).

\section{Discussion}

\subsection{The Effect of Driving Automation on Tactical Self-Regulation}
Our findings on drivers' tactical self-regulation show that drivers adapt their interactions with the center stack touchscreen based on the automation level, vehicle speed, and road curvature (RQ1). Our results show that, drivers perform more touchscreen interactions per sequence with an increasing level of driving automation. Whereas speed influences the number of interactions only slightly, the number of interactions during curved driving decreases by 16\,\%. By breaking down interactions into specific touch gestures, we show that drivers in particular perform more complex gestures like drag and multitouch gestures during automated driving. Drivers also perform significantly less drag and multitouch gestures during curved driving compared to regular tap gestures. Both findings together suggest that drivers adapt their behavior to avoid complex touch gestures in demanding driving situations. They rather engage in such interactions during times of low driving demand. This is in line with the findings of \citet{noble.2021} who found that drivers were more likely to perform high-risk secondary tasks during automated driving sequences.

Concerning drivers' interaction with specific UI elements, we show that during ACC+LCA driving, drivers interact particularly more often with lists or maps compared to other elements like the homebar or AppIcons. A potential explanation for this behavior is that lists and maps are visually more complex and drivers seem to perform these interactions in less demanding driving situations, e.g., with automation enabled or while driving straight. In contrast, the homebar, for example, is easy to access as it is visible on every screen and always located in the same position. The probability of drivers interacting with elements located at the homebar even decreases during ACC+LCA driving compared to manual driving. This could be due to many reasons. One of which may be that during situations of less driving demand, drivers prefer to use interfaces that allow for more control. For example, while it's possible to skip to the next song or radio station using the skip button on the homebar, the media app offers a complete overview of available songs and radio stations. Thus drivers have full control and can choose whatever they prefer. Whereas we observe similar trends for interactions during ACC driving, none of the differences proved to be significant. In our previous work~\citep{ebel.2022}, using parts of the data that we use in this approach, these differences were statistically significant. One reason for this difference could be modifications in the UI software. Since the data is collected from test vehicles, the software is regularly updated so that the UI versions are optimized over time in terms of design, performance, and stability. 

Considering drivers' behavioral adaption of touch gestures and UI elements, it is noticeable that drivers' self-regulation of complex interactions is more sensitive to changes in the driving demand than that of simpler interactions. Meaning that with an increasing driving demand the number of complex interactions decreases faster compared to simpler interactions and vice versa. These findings are in line with previous work~\citep{schneidereit.2017, morgenstern.2020, onate-vega.2020, oviedo-trespalacios.2018,choudhary.2017}, suggesting that drivers tend to perform more demanding tasks in less demanding driving situations. In contrast to related work, which mostly investigates the effects of drivers' tactical self-regulation on a task level, we show that these effects also exist on an interaction level. These new insights can help inform future UI designs for center stack touchscreens.

\subsection{The Effect of Driving Automation on Operational Self-Regulation}
In this study, we show that drivers not only adapt their glance behavior according to the level of driving automation (RQ2), vehicle speed, and road curvature but also show that significant interdependencies between these factors exist (RQ3). These novel findings suggest that drivers extend the margins to which they consider it safe to focus on the center stack touchscreen with an increasing level of driving automation. Even though drivers are supposed to constantly supervise the driving automation~\citep{SAE.2021}, the median glance duration during touchscreen interactions in ACC+LCA driving is 0.59\,s longer than in manual driving. In comparison, \citet{morando.2021} report an average increase of 0.3\,s for glances to the center stack regardless of drivers interacting with the touchscreen. In line with the findings of \citet{noble.2021}, \citet{gaspar.2019}, and \citet{morando.2021}, we also show that drivers are more likely to perform glances longer than two seconds when driving automation is enabled. Whereas \citet{morando.2021} report an increase in the long glance probability toward the center stack touchscreen between manual and level 2 driving of 425\,\%, our results are similar to that of our previous study~\citep{ebel.2022} and suggest an increase of 263\,\%. While the trend is similar, the absolute difference is probably due to differences in the driving environments, the systems under test or the data acquisition.

We also show that during ACC+LCA driving, drivers significantly increase in their mean glance duration toward the center stack touchscreen. This effect is statistically significant across all driving conditions and in line with the model explanations provided by \citet{Ebel.2023}. In contrast, \citet{noble.2021} and \citet{morando.2019} found no significant differences in the mean off-road glance duration for ACC or LCA driving compared to manual driving. There may be two reasons for this: First, the amount of data we leverage in this study is larger. Second, our eye tracker explicitly detects glances toward the center stack touchscreen that we then map to UI interactions. In other studies~\citep{morando.2019, risteska.2021, noble.2021, yang.2021}, authors could not differentiate between general off-path glances, which might still be driving-related, and distraction-related off-path glances. This, inevitably, increases the number of false positives, making it harder to obtain significant results. Considering drivers self-regulation during ACC only driving, drivers increase their glance duration only for straight driving sequences and at speeds between 50--100\,km/h and speeds above 100\,km/h. For all other driving situation the effect is not significant. This suggests that drivers trust the ACC+LCA system to take over at least parts of the driving task in a wide variety of driving situations. On the other hand they only make use of the benefits of the ACC system in relatively controlled driving situations.

\subsection{Limitations and Future Work}

Naturalistic driving studies allow us to observe drivers in their natural and diverse driving environment. Driving simulator studies or test track studies, in contrast, suffer from an \emph{instruction effect} because participants need to perform specific predefined tasks~\citep{carsten.2017}. Furthermore, by leveraging production systems, we collect a large amount of data without the need for, potentially, error-prone manual labeling. However, certain limitations should be considered when interpreting the results.

All cars that contributed to the data collection are company internal test cars. Whereas, they are subject to various testing procedures but also for transfer and leisure rides of employees. Yet, the results of our data analysis do not indicate that specific UI stress tests have been conducted while driving. Furthermore, we argue that even during certain test protocols to evaluate driving-related functions, the incentive to interact with the \acp{IVIS} does not differ from real-world driver behavior. Nonetheless, it is important to note that the software in these test cars is frequently updated and improved. This applies to the UI software as it does to the camera or \ac{ADAS} software. This can lead to changes over time in the way drivers interact with the UI or how they self-regulate their behavior with regard to the driving demand. Compared to our previous work~\citep{ebel.2022}, we can observe differences in the glance and interaction behavior. The differences suggests that drivers' self-regulative behavior is sensitive to small changes in the UI or \ac{ADAS} capabilities. To better understand this effect, similar naturalistic driving studies that compare various \acp{IVIS} and \acp{ADAS} are needed.

Another limitation that is that drivers need to be considered expert users. They are familiar with the cars and additionally obtained a prototype driver's license. Yet, the effect this might have is not clear. Whereas more experienced drivers tend to distribute their visual attention more adequately~\citep{wikman.1998}, \citet{naujoks.2016} report that drivers who are familiar with driving assistance systems are more likely to engage in secondary tasks during assisted driving compared to drivers with no experience. In general, the glance duration distribution is roughly similar to those reported in related studies~\citep{gaspar.2019, morando.2019, noble.2021}.

Due to data privacy regulations, we cannot differentiate between individual drivers. We can only differentiate between different trips and car types. Considering that more than 100 cars, with even more individual drivers, contributed to the data collection, the risk of overfitting to particular drivers is small. However, it is important to consider that only employees contributed to the data collection. For this reason, the results are likely biased toward mid-age drivers.

As we cannot differentiate between individual drivers, we are not able to show personal differences in drivers' self-regulative behavior. However, most of the models fitted (e.g., Model 2 and Model 4 in Appendix \ref{ch:AppendixModels}) in this study have a significantly smaller Marginal $R^2$ compared to the Conditional $R^2$~\citep{Nakagawa.2013}. This indicates that most of the covariance in the data is explained by the fixed and random effects together rather than by the fixed effects only. Even though we only incorporate the trip ($n=10,402$) and car type ($n=138$) as random effects these difference in the Marginal $R^2$ and Conditional $R^2$ suggest that trip-related or personal differences might influence self-regulation. This in line with previous research, but quantifying this effect based on naturalistic data could be the logical next step. The effect of task priority on self-regulation~\citep{Lee.2014} is another factor that is not currently considered, but may provide insights that can aid the design of \acp{IVIS}.

This work could be further improved by incorporating more features that describe the driving demand. Currently, we do not consider environmental factors such as weather and daylight. Speed and curvature may also not be sufficient to distinguish between different driving situations. Low speed and straight driving might be typical for traffic jam behavior (very controlled and easy environment), but also for city driving (very uncontrolled and difficult environment). Including these features could help to provide a more holistic picture of drivers' behavioral adaptations to driving demands.

\section{Conclusion}
We present the first naturalistic driving study to investigate tactical and operational self-regulation of driver interactions with center stack touchscreens. Understanding self-regulation is key to understanding the effects of automation and assistance functions on driver distraction and driving safety. Furthermore, knowledge about self-regulation may help design more user-centered and context-aware \acp{IVIS}. 
The key strengths of our study over the state-of-the-art are two-fold: (1) The large amount of naturalistic data, compared to related approaches~\citep{morando.2019,noble.2021, naujoks.2016}, allows us to investigate drivers' tactical and operational self-regulation in greater detail concerning the driving demand. (2) We evaluate self-regulation specifically during interactions with the center stack touchscreen by combining driving data, UI interactions, touch gestures, and explicit glances toward the center stack touchscreen. That makes this the first naturalistic driving study to show self-regulation based on the analysis of touchscreen interactions.

Our modeling results show that driving automation has a stronger effect on self-regulation than vehicle speed or road curvature. Drivers interact more with the \acs{IVIS} when ACC or ACC+LCA is enabled, use more complex UI elements, and perform more complex touch gestures. Even though driving assistance functions up to level 2 still demand the driver to have full control over the car, we observe 36\% longer glances toward the center stack touchscreen when ACC+LCA is active. 

Further research is needed, but based on the assumption that drivers kept the driving similarly safe throughout all conditions, fixed limits for acceptable demand as reported in the NHTSA Driver Distraction Guidelines~\citep{NHTSA.2014} need to be adjusted according to different levels of driving automation and driving demands.

\section*{Disclosure statement}

The authors declare the following financial interests/personal relationships which may be considered as potential competing interests: Christoph Lingenfelder is an employee of the MBition GmbH which is a subsidiary of Mercedes-Benz. The data used in this work was collected from Mercedes-Benz cars.

\section*{Funding detials}
This project is a part of the PhD research carried out by the first author. The associated project is funded by the MBition GmbH. 

\section*{Data availability statement}
The authors do not have permission to share data.

\bibliographystyle{apacite}
\bibliography{self-regulation_extended.bib}

\begin{thebibliography}{}

\bibitem [\protect \citeauthoryear {%
Bates%
, M{\"a}chler%
, Bolker%
\BCBL {}\ \BBA {} Walker%
}{%
Bates%
\ \protect \BOthers {.}}{%
{\protect \APACyear {2015}}%
}]{%
bates.2015}
\APACinsertmetastar {%
bates.2015}%
\begin{APACrefauthors}%
Bates, D.%
, M{\"a}chler, M.%
, Bolker, B.%
\BCBL {}\ \BBA {} Walker, S.%
\end{APACrefauthors}%
\unskip\
\newblock
\APACrefYearMonthDay{2015}{}{}.
\newblock
{\BBOQ}\APACrefatitle {Fitting {{Linear Mixed-Effects Models Using}}
  {\textbf{Lme4}}} {Fitting {{Linear Mixed-Effects Models Using}}
  {\textbf{lme4}}}.{\BBCQ}
\newblock
\APACjournalVolNumPages{Journal of Statistical Software}{67}{1}{1--48}.
\newblock
\begin{APACrefDOI} \doi{10.18637/jss.v067.i01} \end{APACrefDOI}
\PrintBackRefs{\CurrentBib}

\bibitem [\protect \citeauthoryear {%
Carsten%
\ \protect \BOthers {.}}{%
Carsten%
\ \protect \BOthers {.}}{%
{\protect \APACyear {2017}}%
}]{%
carsten.2017}
\APACinsertmetastar {%
carsten.2017}%
\begin{APACrefauthors}%
Carsten, O.%
, Hibberd, D.%
, B{\"a}rgman, J.%
, Kovaceva, J.%
, Pereira~Cocron, M\BPBI S.%
, Dotzauer, M.%
\BDBL {}Forcolin, F.%
\end{APACrefauthors}%
\unskip\
\newblock
\APACrefYear{2017}.
\newblock
\APACrefbtitle {{{UDRIVE}} Deliverable 43.1, {{Driver Distraction}} and
  {{Inattention}}, of the {{EU FP7 Project UDRIVE}}} {{{UDRIVE}} deliverable
  43.1, {{Driver Distraction}} and {{Inattention}}, of the {{EU FP7 Project
  UDRIVE}}}\ (\PrintOrdinal{First}\ \BEd).
\newblock
\APACaddressPublisher{{BE}}{{UDRIVE Consortium}}.
\PrintBackRefs{\CurrentBib}

\bibitem [\protect \citeauthoryear {%
Choudhary%
\ \BBA {} Velaga%
}{%
Choudhary%
\ \BBA {} Velaga%
}{%
{\protect \APACyear {2017}}%
}]{%
choudhary.2017}
\APACinsertmetastar {%
choudhary.2017}%
\begin{APACrefauthors}%
Choudhary, P.%
\BCBT {}\ \BBA {} Velaga, N\BPBI R.%
\end{APACrefauthors}%
\unskip\
\newblock
\APACrefYearMonthDay{2017}{{\APACmonth{09}}}{}.
\newblock
{\BBOQ}\APACrefatitle {Mobile Phone Use during Driving: {{Effects}} on Speed
  and Effectiveness of Driver Compensatory Behaviour} {Mobile phone use during
  driving: {{Effects}} on speed and effectiveness of driver compensatory
  behaviour}.{\BBCQ}
\newblock
\APACjournalVolNumPages{Accident Analysis \& Prevention}{106}{}{370--378}.
\newblock
\begin{APACrefDOI} \doi{10.1016/j.aap.2017.06.021} \end{APACrefDOI}
\PrintBackRefs{\CurrentBib}

\bibitem [\protect \citeauthoryear {%
Christoph%
, Wesseling%
\BCBL {}\ \BBA {} {van Nes}%
}{%
Christoph%
\ \protect \BOthers {.}}{%
{\protect \APACyear {2019}}%
}]{%
christoph.2019}
\APACinsertmetastar {%
christoph.2019}%
\begin{APACrefauthors}%
Christoph, M.%
, Wesseling, S.%
\BCBL {}\ \BBA {} {van Nes}, N.%
\end{APACrefauthors}%
\unskip\
\newblock
\APACrefYearMonthDay{2019}{{\APACmonth{10}}}{}.
\newblock
{\BBOQ}\APACrefatitle {Self-Regulation of Drivers' Mobile Phone Use: {{The}}
  Influence of Driving Context} {Self-regulation of drivers' mobile phone use:
  {{The}} influence of driving context}.{\BBCQ}
\newblock
\APACjournalVolNumPages{Transportation Research Part F: Traffic Psychology and
  Behaviour}{66}{}{262--272}.
\newblock
\begin{APACrefDOI} \doi{10.1016/j.trf.2019.09.012} \end{APACrefDOI}
\PrintBackRefs{\CurrentBib}

\bibitem [\protect \citeauthoryear {%
DeGuzman%
\ \BBA {} Donmez%
}{%
DeGuzman%
\ \BBA {} Donmez%
}{%
{\protect \APACyear {2021}}%
}]{%
deguzman.2021}
\APACinsertmetastar {%
deguzman.2021}%
\begin{APACrefauthors}%
DeGuzman, C\BPBI A.%
\BCBT {}\ \BBA {} Donmez, B.%
\end{APACrefauthors}%
\unskip\
\newblock
\APACrefYearMonthDay{2021}{{\APACmonth{06}}}{}.
\newblock
{\BBOQ}\APACrefatitle {Knowledge of and Trust in Advanced Driver Assistance
  Systems} {Knowledge of and trust in advanced driver assistance
  systems}.{\BBCQ}
\newblock
\APACjournalVolNumPages{Accident Analysis \& Prevention}{156}{}{106121}.
\newblock
\begin{APACrefDOI} \doi{10.1016/j.aap.2021.106121} \end{APACrefDOI}
\PrintBackRefs{\CurrentBib}

\bibitem [\protect \citeauthoryear {%
Dingus%
\ \protect \BOthers {.}}{%
Dingus%
\ \protect \BOthers {.}}{%
{\protect \APACyear {2016}}%
}]{%
dingus.2016}
\APACinsertmetastar {%
dingus.2016}%
\begin{APACrefauthors}%
Dingus, T\BPBI A.%
, Guo, F.%
, Lee, S.%
, Antin, J\BPBI F.%
, Perez, M.%
, {Buchanan-King}, M.%
\BCBL {}\ \BBA {} Hankey, J.%
\end{APACrefauthors}%
\unskip\
\newblock
\APACrefYearMonthDay{2016}{{\APACmonth{03}}}{}.
\newblock
{\BBOQ}\APACrefatitle {Driver Crash Risk Factors and Prevalence Evaluation
  Using Naturalistic Driving Data} {Driver crash risk factors and prevalence
  evaluation using naturalistic driving data}.{\BBCQ}
\newblock
\APACjournalVolNumPages{Proceedings of the National Academy of
  Sciences}{113}{10}{2636--2641}.
\newblock
\begin{APACrefDOI} \doi{10.1073/pnas.1513271113} \end{APACrefDOI}
\PrintBackRefs{\CurrentBib}

\bibitem [\protect \citeauthoryear {%
Dunn%
, Dingus%
\BCBL {}\ \BBA {} Soccolich%
}{%
Dunn%
\ \protect \BOthers {.}}{%
{\protect \APACyear {2020}}%
}]{%
dunn.2020}
\APACinsertmetastar {%
dunn.2020}%
\begin{APACrefauthors}%
Dunn, N.%
, Dingus, T.%
\BCBL {}\ \BBA {} Soccolich, S.%
\end{APACrefauthors}%
\unskip\
\newblock
\APACrefYearMonthDay{2020}{{\APACmonth{02}}}{}.
\newblock
{\BBOQ}\APACrefatitle {What's in a Name? {{Drivers}}' Perceptions of the Use of
  Five {{SAE Level}} 2 Driving Automation Systems} {What's in a name?
  {{Drivers}}' perceptions of the use of five {{SAE Level}} 2 driving
  automation systems}.{\BBCQ}
\newblock
\APACjournalVolNumPages{Journal of Safety Research}{72}{}{145--151}.
\newblock
\begin{APACrefDOI} \doi{10.1016/j.jsr.2019.11.005} \end{APACrefDOI}
\PrintBackRefs{\CurrentBib}

\bibitem [\protect \citeauthoryear {%
Ebel%
, Berger%
, Lingenfelder%
\BCBL {}\ \BBA {} Vogelsang%
}{%
Ebel%
\ \protect \BOthers {.}}{%
{\protect \APACyear {2022}}%
}]{%
ebel.2022}
\APACinsertmetastar {%
ebel.2022}%
\begin{APACrefauthors}%
Ebel, P.%
, Berger, M.%
, Lingenfelder, C.%
\BCBL {}\ \BBA {} Vogelsang, A.%
\end{APACrefauthors}%
\unskip\
\newblock
\APACrefYearMonthDay{2022}{{\APACmonth{09}}}{}.
\newblock
{\BBOQ}\APACrefatitle {How {{Do Drivers Self-Regulate}} Their {{Secondary Task
  Engagements}}? {{The Effect}} of {{Driving Automation}} on {{Touchscreen
  Interactions}} and {{Glance Behavior}}} {How {{Do Drivers Self-Regulate}}
  their {{Secondary Task Engagements}}? {{The Effect}} of {{Driving
  Automation}} on {{Touchscreen Interactions}} and {{Glance Behavior}}}.{\BBCQ}
\newblock
\BIn{} \APACrefbtitle {Proceedings of the 14th {{International Conference}} on
  {{Automotive User Interfaces}} and {{Interactive Vehicular Applications}}}
  {Proceedings of the 14th {{International Conference}} on {{Automotive User
  Interfaces}} and {{Interactive Vehicular Applications}}}\ (\BPGS\ 263--273).
\newblock
\APACaddressPublisher{{Seoul Republic of Korea}}{{ACM}}.
\newblock
\begin{APACrefDOI} \doi{10.1145/3543174.3545173} \end{APACrefDOI}
\PrintBackRefs{\CurrentBib}

\bibitem [\protect \citeauthoryear {%
Ebel%
, Lingenfelder%
\BCBL {}\ \BBA {} Vogelsang%
}{%
Ebel%
, Lingenfelder%
\BCBL {}\ \BBA {} Vogelsang%
}{%
{\protect \APACyear {2021}}%
}]{%
ebel.2021a}
\APACinsertmetastar {%
ebel.2021a}%
\begin{APACrefauthors}%
Ebel, P.%
, Lingenfelder, C.%
\BCBL {}\ \BBA {} Vogelsang, A.%
\end{APACrefauthors}%
\unskip\
\newblock
\APACrefYearMonthDay{2021}{{\APACmonth{09}}}{}.
\newblock
{\BBOQ}\APACrefatitle {Visualizing {{Event Sequence Data}} for {{User Behavior
  Evaluation}} of {{In-Vehicle Information Systems}}} {Visualizing {{Event
  Sequence Data}} for {{User Behavior Evaluation}} of {{In-Vehicle Information
  Systems}}}.{\BBCQ}
\newblock
\BIn{} \APACrefbtitle {13th {{International Conference}} on {{Automotive User
  Interfaces}} and {{Interactive Vehicular Applications}}} {13th
  {{International Conference}} on {{Automotive User Interfaces}} and
  {{Interactive Vehicular Applications}}}\ (\BPGS\ 219--229).
\newblock
\APACaddressPublisher{{Leeds United Kingdom}}{{ACM}}.
\newblock
\begin{APACrefDOI} \doi{10.1145/3409118.3475140} \end{APACrefDOI}
\PrintBackRefs{\CurrentBib}

\bibitem [\protect \citeauthoryear {%
Ebel%
, Lingenfelder%
\BCBL {}\ \BBA {} Vogelsang%
}{%
Ebel%
\ \protect \BOthers {.}}{%
{\protect \APACyear {2023}}%
}]{%
Ebel.2023}
\APACinsertmetastar {%
Ebel.2023}%
\begin{APACrefauthors}%
Ebel, P.%
, Lingenfelder, C.%
\BCBL {}\ \BBA {} Vogelsang, A.%
\end{APACrefauthors}%
\unskip\
\newblock
\APACrefYearMonthDay{2023}{{\APACmonth{04}}}{}.
\newblock
{\BBOQ}\APACrefatitle {On the Forces of Driver Distraction: {{Explainable}}
  Predictions for the Visual Demand of in-Vehicle Touchscreen Interactions} {On
  the forces of driver distraction: {{Explainable}} predictions for the visual
  demand of in-vehicle touchscreen interactions}.{\BBCQ}
\newblock
\APACjournalVolNumPages{Accident Analysis \& Prevention}{183}{}{106956}.
\newblock
\begin{APACrefDOI} \doi{10.1016/j.aap.2023.106956} \end{APACrefDOI}
\PrintBackRefs{\CurrentBib}

\bibitem [\protect \citeauthoryear {%
Ebel%
, Orlovska%
\BCBL {}\ \protect \BOthers {.}}{%
Ebel%
, Orlovska%
\BCBL {}\ \protect \BOthers {.}}{%
{\protect \APACyear {2021}}%
}]{%
ebel.2021}
\APACinsertmetastar {%
ebel.2021}%
\begin{APACrefauthors}%
Ebel, P.%
, Orlovska, J.%
, H{\"u}nemeyer, S.%
, Wickman, C.%
, Vogelsang, A.%
\BCBL {}\ \BBA {} S{\"o}derberg, R.%
\end{APACrefauthors}%
\unskip\
\newblock
\APACrefYearMonthDay{2021}{{\APACmonth{09}}}{}.
\newblock
{\BBOQ}\APACrefatitle {Automotive {{UX}} Design and Data-Driven Development:
  {{Narrowing}} the Gap to Support Practitioners} {Automotive {{UX}} design and
  data-driven development: {{Narrowing}} the gap to support
  practitioners}.{\BBCQ}
\newblock
\APACjournalVolNumPages{Transportation Research Interdisciplinary
  Perspectives}{11}{}{100455}.
\newblock
\begin{APACrefDOI} \doi{10.1016/j.trip.2021.100455} \end{APACrefDOI}
\PrintBackRefs{\CurrentBib}

\bibitem [\protect \citeauthoryear {%
Ervin%
\ \protect \BOthers {.}}{%
Ervin%
\ \protect \BOthers {.}}{%
{\protect \APACyear {2005}}%
}]{%
Ervin.2005}
\APACinsertmetastar {%
Ervin.2005}%
\begin{APACrefauthors}%
Ervin, R.%
, Sayer, J.%
, LeBlanc, D.%
, Bogard, S.%
, Mefford, M.%
, Hagan, Z.%
\BDBL {}Winkler, C.%
\end{APACrefauthors}%
\unskip\
\newblock
\APACrefYearMonthDay{2005}{}{}.
\newblock
\APACrefbtitle {Automotive Collision Avoidance System Field Operational Test
  Report: Methodology and Results} {Automotive collision avoidance system field
  operational test report: Methodology and results}\ \APACbVolEdTR{}{\BTR{}\
  \BNUM\ HS-809 900}.
\newblock
\APACaddressInstitution{}{{National Highway Traffic Safety Administration}}.
\PrintBackRefs{\CurrentBib}

\bibitem [\protect \citeauthoryear {%
Faber%
\ \protect \BOthers {.}}{%
Faber%
\ \protect \BOthers {.}}{%
{\protect \APACyear {2012}}%
}]{%
Faber.2012}
\APACinsertmetastar {%
Faber.2012}%
\begin{APACrefauthors}%
Faber, F.%
, Jonkers, E.%
, {van Noort}, M.%
, Benmimoun, M.%
, Andreas, P.%
, Metz, B.%
\BDBL {}Malta, L.%
\end{APACrefauthors}%
\unskip\
\newblock
\APACrefYearMonthDay{2012}{}{}.
\newblock
\APACrefbtitle {{{EuroFOT}} Deliverable 6.4 - Final Results: {{Impacts}} on
  Traffic Safety} {{{EuroFOT}} deliverable 6.4 - final results: {{Impacts}} on
  traffic safety}\ \APACbVolEdTR {}{Project {{Report}}}.
\newblock
\APACaddressInstitution{}{{ERTICO - ITS Europe}}.
\PrintBackRefs{\CurrentBib}

\bibitem [\protect \citeauthoryear {%
Gaspar%
\ \BBA {} Carney%
}{%
Gaspar%
\ \BBA {} Carney%
}{%
{\protect \APACyear {2019}}%
}]{%
gaspar.2019}
\APACinsertmetastar {%
gaspar.2019}%
\begin{APACrefauthors}%
Gaspar, J.%
\BCBT {}\ \BBA {} Carney, C.%
\end{APACrefauthors}%
\unskip\
\newblock
\APACrefYearMonthDay{2019}{{\APACmonth{12}}}{}.
\newblock
{\BBOQ}\APACrefatitle {The {{Effect}} of {{Partial Automation}} on {{Driver
  Attention}}: {{A Naturalistic Driving Study}}} {The {{Effect}} of {{Partial
  Automation}} on {{Driver Attention}}: {{A Naturalistic Driving
  Study}}}.{\BBCQ}
\newblock
\APACjournalVolNumPages{Human Factors: The Journal of the Human Factors and
  Ergonomics Society}{61}{8}{1261--1276}.
\newblock
\begin{APACrefDOI} \doi{10.1177/0018720819836310} \end{APACrefDOI}
\PrintBackRefs{\CurrentBib}

\bibitem [\protect \citeauthoryear {%
Hancox%
, Richardson%
\BCBL {}\ \BBA {} Morris%
}{%
Hancox%
\ \protect \BOthers {.}}{%
{\protect \APACyear {2013}}%
}]{%
hancox.2013}
\APACinsertmetastar {%
hancox.2013}%
\begin{APACrefauthors}%
Hancox, G.%
, Richardson, J.%
\BCBL {}\ \BBA {} Morris, A.%
\end{APACrefauthors}%
\unskip\
\newblock
\APACrefYearMonthDay{2013}{{\APACmonth{06}}}{}.
\newblock
{\BBOQ}\APACrefatitle {Drivers' Willingness to Engage with Their Mobile Phone:
  The Influence of Phone Function and Road Demand} {Drivers' willingness to
  engage with their mobile phone: The influence of phone function and road
  demand}.{\BBCQ}
\newblock
\APACjournalVolNumPages{IET Intelligent Transport Systems}{7}{2}{215--222}.
\newblock
\begin{APACrefDOI} \doi{10.1049/iet-its.2012.0133} \end{APACrefDOI}
\PrintBackRefs{\CurrentBib}

\bibitem [\protect \citeauthoryear {%
Hlavac%
}{%
Hlavac%
}{%
{\protect \APACyear {2022}}%
}]{%
hlavac.2022}
\APACinsertmetastar {%
hlavac.2022}%
\begin{APACrefauthors}%
Hlavac, M.%
\end{APACrefauthors}%
\unskip\
\newblock
\APACrefYearMonthDay{2022}{}{}.
\newblock
\APACrefbtitle {Stargazer: {{Well-formatted}} Regression and Summary Statistics
  Tables} {Stargazer: {{Well-formatted}} regression and summary statistics
  tables}\ [Manual].
\newblock
\APACaddressPublisher{{Bratislava, Slovakia}}{}.
\PrintBackRefs{\CurrentBib}

\bibitem [\protect \citeauthoryear {%
Horrey%
\ \BBA {} Lesch%
}{%
Horrey%
\ \BBA {} Lesch%
}{%
{\protect \APACyear {2009}}%
}]{%
horrey.2009}
\APACinsertmetastar {%
horrey.2009}%
\begin{APACrefauthors}%
Horrey, W\BPBI J.%
\BCBT {}\ \BBA {} Lesch, M\BPBI F.%
\end{APACrefauthors}%
\unskip\
\newblock
\APACrefYearMonthDay{2009}{{\APACmonth{01}}}{}.
\newblock
{\BBOQ}\APACrefatitle {Driver-Initiated Distractions: {{Examining}} Strategic
  Adaptation for in-Vehicle Task Initiation} {Driver-initiated distractions:
  {{Examining}} strategic adaptation for in-vehicle task initiation}.{\BBCQ}
\newblock
\APACjournalVolNumPages{Accident Analysis \& Prevention}{41}{1}{115--122}.
\newblock
\begin{APACrefDOI} \doi{10.1016/j.aap.2008.10.008} \end{APACrefDOI}
\PrintBackRefs{\CurrentBib}

\bibitem [\protect \citeauthoryear {%
Hox%
}{%
Hox%
}{%
{\protect \APACyear {1998}}%
}]{%
hox.1998}
\APACinsertmetastar {%
hox.1998}%
\begin{APACrefauthors}%
Hox, J.%
\end{APACrefauthors}%
\unskip\
\newblock
\APACrefYearMonthDay{1998}{}{}.
\newblock
{\BBOQ}\APACrefatitle {Multilevel {{Modeling}}: {{When}} and {{Why}}}
  {Multilevel {{Modeling}}: {{When}} and {{Why}}}.{\BBCQ}
\newblock
\BIn{} H\BPBI H.~Bock, O.~Opitz, M.~Schader, I.~Balderjahn, R.~Mathar\BCBL {}\
  \BBA {} M.~Schader\ (\BEDS), \APACrefbtitle {Classification, {{Data
  Analysis}}, and {{Data Highways}}} {Classification, {{Data Analysis}}, and
  {{Data Highways}}}\ (\BPGS\ 147--154).
\newblock
\APACaddressPublisher{{Berlin, Heidelberg}}{{Springer Berlin Heidelberg}}.
\newblock
\begin{APACrefDOI} \doi{10.1007/978-3-642-72087-1_17} \end{APACrefDOI}
\PrintBackRefs{\CurrentBib}

\bibitem [\protect \citeauthoryear {%
Hutchinson%
, White%
, Martin%
, Reichert%
\BCBL {}\ \BBA {} Frey%
}{%
Hutchinson%
\ \protect \BOthers {.}}{%
{\protect \APACyear {1989}}%
}]{%
hutchinson.1989}
\APACinsertmetastar {%
hutchinson.1989}%
\begin{APACrefauthors}%
Hutchinson, T.%
, White, K.%
, Martin, W.%
, Reichert, K.%
\BCBL {}\ \BBA {} Frey, L.%
\end{APACrefauthors}%
\unskip\
\newblock
\APACrefYearMonthDay{1989}{}{}.
\newblock
{\BBOQ}\APACrefatitle {Human-Computer Interaction Using Eye-Gaze Input}
  {Human-computer interaction using eye-gaze input}.{\BBCQ}
\newblock
\APACjournalVolNumPages{IEEE Transactions on Systems, Man, and
  Cybernetics}{19}{6}{1527--1534}.
\newblock
\begin{APACrefDOI} \doi{10.1109/21.44068} \end{APACrefDOI}
\PrintBackRefs{\CurrentBib}

\bibitem [\protect \citeauthoryear {%
Ismaeel%
, Hibberd%
, Carsten%
\BCBL {}\ \BBA {} Jamson%
}{%
Ismaeel%
\ \protect \BOthers {.}}{%
{\protect \APACyear {2020}}%
}]{%
ismaeel.2020}
\APACinsertmetastar {%
ismaeel.2020}%
\begin{APACrefauthors}%
Ismaeel, R.%
, Hibberd, D.%
, Carsten, O.%
\BCBL {}\ \BBA {} Jamson, S.%
\end{APACrefauthors}%
\unskip\
\newblock
\APACrefYearMonthDay{2020}{{\APACmonth{03}}}{}.
\newblock
{\BBOQ}\APACrefatitle {Do Drivers Self-Regulate Their Engagement in Secondary
  Tasks at Intersections? {{An}} Examination Based on Naturalistic Driving
  Data} {Do drivers self-regulate their engagement in secondary tasks at
  intersections? {{An}} examination based on naturalistic driving data}.{\BBCQ}
\newblock
\APACjournalVolNumPages{Accident Analysis \& Prevention}{137}{}{105464}.
\newblock
\begin{APACrefDOI} \doi{10.1016/j.aap.2020.105464} \end{APACrefDOI}
\PrintBackRefs{\CurrentBib}

\bibitem [\protect \citeauthoryear {%
{ISO/TC 22/SC 39 Ergonomics}%
}{%
{ISO/TC 22/SC 39 Ergonomics}%
}{%
{\protect \APACyear {2020}}%
}]{%
ISO15007}
\APACinsertmetastar {%
ISO15007}%
\begin{APACrefauthors}%
{ISO/TC 22/SC 39 Ergonomics}.%
\end{APACrefauthors}%
\unskip\
\newblock
\APACrefYearMonthDay{2020}{}{}.
\newblock
\APACrefbtitle {Road Vehicles \textemdash{} {{Measurement}} and Analysis of
  Driver Visual Behaviour with Respect to Transport Information and Control
  Systems} {Road vehicles \textemdash{} {{Measurement}} and analysis of driver
  visual behaviour with respect to transport information and control systems}\
  \APACbVolEdTR{}{\BTR{}}.
\newblock
\APACaddressInstitution{}{{International Organization for Standardization}}.
\PrintBackRefs{\CurrentBib}

\bibitem [\protect \citeauthoryear {%
Klauer%
, Dingus%
, Neale%
, Sudweeks%
\BCBL {}\ \BBA {} Ramsey%
}{%
Klauer%
\ \protect \BOthers {.}}{%
{\protect \APACyear {2006}}%
}]{%
Klauer.2006}
\APACinsertmetastar {%
Klauer.2006}%
\begin{APACrefauthors}%
Klauer, S.%
, Dingus, T.%
, Neale, T.%
, Sudweeks, J.%
\BCBL {}\ \BBA {} Ramsey, D.%
\end{APACrefauthors}%
\unskip\
\newblock
\APACrefYearMonthDay{2006}{{\APACmonth{01}}}{}.
\newblock
\APACrefbtitle {The Impact of Driver Inattention on Near-Crash/Crash Risk:
  {{An}} Analysis Using the 100-{{Car}} Naturalistic Driving Study Data} {The
  impact of driver inattention on near-crash/crash risk: {{An}} analysis using
  the 100-{{Car}} naturalistic driving study data}\
  \APACbVolEdTR{\BVOL~594}{\BTR{}}.
\newblock
\APACaddressInstitution{{3500 Transportation Research Plaza (0536) Blacksburg,
  Virginia 24061}}{{U.S. Department of Transportation, National Highway Traffic
  Safety Administration / Virginia Tech Transportation Institute}}.
\PrintBackRefs{\CurrentBib}

\bibitem [\protect \citeauthoryear {%
Kuznetsova%
, Brockhoff%
\BCBL {}\ \BBA {} Christensen%
}{%
Kuznetsova%
\ \protect \BOthers {.}}{%
{\protect \APACyear {2017}}%
}]{%
kuznetsova.2017}
\APACinsertmetastar {%
kuznetsova.2017}%
\begin{APACrefauthors}%
Kuznetsova, A.%
, Brockhoff, P\BPBI B.%
\BCBL {}\ \BBA {} Christensen, R\BPBI H\BPBI B.%
\end{APACrefauthors}%
\unskip\
\newblock
\APACrefYearMonthDay{2017}{}{}.
\newblock
{\BBOQ}\APACrefatitle {{{{\textbf{lmerTest}}}} {{Package}}: {{Tests}} in
  {{Linear Mixed Effects Models}}} {{{{\textbf{lmerTest}}}} {{Package}}:
  {{Tests}} in {{Linear Mixed Effects Models}}}.{\BBCQ}
\newblock
\APACjournalVolNumPages{Journal of Statistical Software}{82}{13}{1--26}.
\newblock
\begin{APACrefDOI} \doi{10.18637/jss.v082.i13} \end{APACrefDOI}
\PrintBackRefs{\CurrentBib}

\bibitem [\protect \citeauthoryear {%
J.~Lee%
, Young%
\BCBL {}\ \BBA {} Regan%
}{%
J.~Lee%
\ \protect \BOthers {.}}{%
{\protect \APACyear {2008}}%
}]{%
lee.2008}
\APACinsertmetastar {%
lee.2008}%
\begin{APACrefauthors}%
Lee, J.%
, Young, K.%
\BCBL {}\ \BBA {} Regan, M.%
\end{APACrefauthors}%
\unskip\
\newblock
\APACrefYearMonthDay{2008}{{\APACmonth{10}}}{}.
\newblock
{\BBOQ}\APACrefatitle {Defining {{Driver Distraction}}} {Defining {{Driver
  Distraction}}}.{\BBCQ}
\newblock
\BIn{} M.~Regan, J.~Lee\BCBL {}\ \BBA {} K.~Young\ (\BEDS), \APACrefbtitle
  {Driver {{Distraction}}} {Driver {{Distraction}}}\ (\BPGS\ 31--40).
\newblock
\APACaddressPublisher{}{{CRC Press}}.
\newblock
\begin{APACrefDOI} \doi{10.1201/9781420007497.ch3} \end{APACrefDOI}
\PrintBackRefs{\CurrentBib}

\bibitem [\protect \citeauthoryear {%
J\BPBI D.~Lee%
}{%
J\BPBI D.~Lee%
}{%
{\protect \APACyear {2014}}%
}]{%
Lee.2014}
\APACinsertmetastar {%
Lee.2014}%
\begin{APACrefauthors}%
Lee, J\BPBI D.%
\end{APACrefauthors}%
\unskip\
\newblock
\APACrefYearMonthDay{2014}{}{}.
\newblock
{\BBOQ}\APACrefatitle {Dynamics of {{Driver Distraction}}: {{The}} Process of
  Engaging and Disengaging} {Dynamics of {{Driver Distraction}}: {{The}}
  process of engaging and disengaging}.{\BBCQ}
\newblock
\APACjournalVolNumPages{Annals of Advances in Automotive Medicine. Association
  for the Advancement of Automotive Medicine. Annual Scientific
  Conference}{58}{}{24--32}.
\PrintBackRefs{\CurrentBib}

\bibitem [\protect \citeauthoryear {%
Lenth%
}{%
Lenth%
}{%
{\protect \APACyear {2022}}%
}]{%
lenth.2022}
\APACinsertmetastar {%
lenth.2022}%
\begin{APACrefauthors}%
Lenth, R\BPBI V.%
\end{APACrefauthors}%
\unskip\
\newblock
\APACrefYearMonthDay{2022}{}{}.
\newblock
\APACrefbtitle {Emmeans: {{Estimated}} Marginal Means, Aka Least-Squares
  Means.} {Emmeans: {{Estimated}} marginal means, aka least-squares means.}
\PrintBackRefs{\CurrentBib}

\bibitem [\protect \citeauthoryear {%
Liang%
, Horrey%
\BCBL {}\ \BBA {} Hoffman%
}{%
Liang%
\ \protect \BOthers {.}}{%
{\protect \APACyear {2015}}%
}]{%
liang.2015}
\APACinsertmetastar {%
liang.2015}%
\begin{APACrefauthors}%
Liang, Y.%
, Horrey, W\BPBI J.%
\BCBL {}\ \BBA {} Hoffman, J\BPBI D.%
\end{APACrefauthors}%
\unskip\
\newblock
\APACrefYearMonthDay{2015}{{\APACmonth{03}}}{}.
\newblock
{\BBOQ}\APACrefatitle {Reading {{Text While Driving}}: {{Understanding
  Drivers}}' {{Strategic}} and {{Tactical Adaptation}} to {{Distraction}}}
  {Reading {{Text While Driving}}: {{Understanding Drivers}}' {{Strategic}} and
  {{Tactical Adaptation}} to {{Distraction}}}.{\BBCQ}
\newblock
\APACjournalVolNumPages{Human Factors: The Journal of the Human Factors and
  Ergonomics Society}{57}{2}{347--359}.
\newblock
\begin{APACrefDOI} \doi{10.1177/0018720814542974} \end{APACrefDOI}
\PrintBackRefs{\CurrentBib}

\bibitem [\protect \citeauthoryear {%
Lin%
, Liu%
, Ma%
, Zhang%
\BCBL {}\ \BBA {} Zhang%
}{%
Lin%
\ \protect \BOthers {.}}{%
{\protect \APACyear {2019}}%
}]{%
lin.2019}
\APACinsertmetastar {%
lin.2019}%
\begin{APACrefauthors}%
Lin, R.%
, Liu, N.%
, Ma, L.%
, Zhang, T.%
\BCBL {}\ \BBA {} Zhang, W.%
\end{APACrefauthors}%
\unskip\
\newblock
\APACrefYearMonthDay{2019}{{\APACmonth{07}}}{}.
\newblock
{\BBOQ}\APACrefatitle {Exploring the Self-Regulation of Secondary Task
  Engagement in the Context of Partially Automated Driving: {{A}} Pilot Study}
  {Exploring the self-regulation of secondary task engagement in the context of
  partially automated driving: {{A}} pilot study}.{\BBCQ}
\newblock
\APACjournalVolNumPages{Transportation Research Part F: Traffic Psychology and
  Behaviour}{64}{}{147--160}.
\newblock
\begin{APACrefDOI} \doi{10.1016/j.trf.2019.05.005} \end{APACrefDOI}
\PrintBackRefs{\CurrentBib}

\bibitem [\protect \citeauthoryear {%
Luke%
}{%
Luke%
}{%
{\protect \APACyear {2017}}%
}]{%
luke.2017}
\APACinsertmetastar {%
luke.2017}%
\begin{APACrefauthors}%
Luke, S\BPBI G.%
\end{APACrefauthors}%
\unskip\
\newblock
\APACrefYearMonthDay{2017}{{\APACmonth{08}}}{}.
\newblock
{\BBOQ}\APACrefatitle {Evaluating Significance in Linear Mixed-Effects Models
  in {{R}}} {Evaluating significance in linear mixed-effects models in
  {{R}}}.{\BBCQ}
\newblock
\APACjournalVolNumPages{Behavior Research Methods}{49}{4}{1494--1502}.
\newblock
\begin{APACrefDOI} \doi{10.3758/s13428-016-0809-y} \end{APACrefDOI}
\PrintBackRefs{\CurrentBib}

\bibitem [\protect \citeauthoryear {%
Magezi%
}{%
Magezi%
}{%
{\protect \APACyear {2015}}%
}]{%
magezi.2015}
\APACinsertmetastar {%
magezi.2015}%
\begin{APACrefauthors}%
Magezi, D\BPBI A.%
\end{APACrefauthors}%
\unskip\
\newblock
\APACrefYearMonthDay{2015}{{\APACmonth{01}}}{}.
\newblock
{\BBOQ}\APACrefatitle {Linear Mixed-Effects Models for within-Participant
  Psychology Experiments: An Introductory Tutorial and Free, Graphical User
  Interface ({{LMMgui}})} {Linear mixed-effects models for within-participant
  psychology experiments: An introductory tutorial and free, graphical user
  interface ({{LMMgui}})}.{\BBCQ}
\newblock
\APACjournalVolNumPages{Frontiers in Psychology}{6}{}{2}.
\newblock
\begin{APACrefDOI} \doi{10.3389/fpsyg.2015.00002} \end{APACrefDOI}
\PrintBackRefs{\CurrentBib}

\bibitem [\protect \citeauthoryear {%
{Mercedes-Benz Group AG}%
}{%
{Mercedes-Benz Group AG}%
}{%
{\protect \APACyear {2023}}%
}]{%
Mercedes.2023}
\APACinsertmetastar {%
Mercedes.2023}%
\begin{APACrefauthors}%
{Mercedes-Benz Group AG}.%
\end{APACrefauthors}%
\unskip\
\newblock
\APACrefYearMonthDay{2023}{{\APACmonth{01}}}{}.
\newblock
\APACrefbtitle {Mercedes-{{Benz Reveals New Charging Network}} and {{Tech
  Updates}} at {{CES}} 2023.} {Mercedes-{{Benz Reveals New Charging Network}}
  and {{Tech Updates}} at {{CES}} 2023.}
\newblock
\APAChowpublished
  {\url{https://media.mercedes-benz.com/article/81dc6cca-71a8-4199-bfa9-8484e4e80678}}.
\PrintBackRefs{\CurrentBib}

\bibitem [\protect \citeauthoryear {%
Merchant%
}{%
Merchant%
}{%
{\protect \APACyear {1967}}%
}]{%
merchant.1967}
\APACinsertmetastar {%
merchant.1967}%
\begin{APACrefauthors}%
Merchant, J.%
\end{APACrefauthors}%
\unskip\
\newblock
\APACrefYearMonthDay{1967}{}{}.
\newblock
\APACrefbtitle {The Oculometer} {The oculometer}\ \APACbVolEdTR{}{\BTR{}\
  \BNUM\ NASA-CR-805}.
\newblock
\APACaddressInstitution{}{{National Aeronautics and Space Administration}}.
\PrintBackRefs{\CurrentBib}

\bibitem [\protect \citeauthoryear {%
Michon%
}{%
Michon%
}{%
{\protect \APACyear {1985}}%
}]{%
michon.1985}
\APACinsertmetastar {%
michon.1985}%
\begin{APACrefauthors}%
Michon, J\BPBI A.%
\end{APACrefauthors}%
\unskip\
\newblock
\APACrefYearMonthDay{1985}{}{}.
\newblock
{\BBOQ}\APACrefatitle {A {{Critical View}} of {{Driver Behavior Models}}:
  {{What Do We Know}}, {{What Should We Do}}?} {A {{Critical View}} of {{Driver
  Behavior Models}}: {{What Do We Know}}, {{What Should We Do}}?}{\BBCQ}
\newblock
\BIn{} L.~Evans\ \BBA {} R\BPBI C.~Schwing\ (\BEDS), \APACrefbtitle {Human
  {{Behavior}} and {{Traffic Safety}}} {Human {{Behavior}} and {{Traffic
  Safety}}}\ (\BPGS\ 485--524).
\newblock
\APACaddressPublisher{{Boston, MA}}{{Springer US}}.
\newblock
\begin{APACrefDOI} \doi{10.1007/978-1-4613-2173-6_19} \end{APACrefDOI}
\PrintBackRefs{\CurrentBib}

\bibitem [\protect \citeauthoryear {%
Morando%
, Gershon%
, Mehler%
\BCBL {}\ \BBA {} Reimer%
}{%
Morando%
\ \protect \BOthers {.}}{%
{\protect \APACyear {2021}}%
}]{%
morando.2021}
\APACinsertmetastar {%
morando.2021}%
\begin{APACrefauthors}%
Morando, A.%
, Gershon, P.%
, Mehler, B.%
\BCBL {}\ \BBA {} Reimer, B.%
\end{APACrefauthors}%
\unskip\
\newblock
\APACrefYearMonthDay{2021}{{\APACmonth{10}}}{}.
\newblock
{\BBOQ}\APACrefatitle {A Model for Naturalistic Glance Behavior around {{Tesla
  Autopilot}} Disengagements} {A model for naturalistic glance behavior around
  {{Tesla Autopilot}} disengagements}.{\BBCQ}
\newblock
\APACjournalVolNumPages{Accident Analysis \& Prevention}{161}{}{106348}.
\newblock
\begin{APACrefDOI} \doi{10.1016/j.aap.2021.106348} \end{APACrefDOI}
\PrintBackRefs{\CurrentBib}

\bibitem [\protect \citeauthoryear {%
Morando%
, Victor%
\BCBL {}\ \BBA {} Dozza%
}{%
Morando%
\ \protect \BOthers {.}}{%
{\protect \APACyear {2019}}%
}]{%
morando.2019}
\APACinsertmetastar {%
morando.2019}%
\begin{APACrefauthors}%
Morando, A.%
, Victor, T.%
\BCBL {}\ \BBA {} Dozza, M.%
\end{APACrefauthors}%
\unskip\
\newblock
\APACrefYearMonthDay{2019}{{\APACmonth{08}}}{}.
\newblock
{\BBOQ}\APACrefatitle {A {{Reference Model}} for {{Driver Attention}} in
  {{Automation}}: {{Glance Behavior Changes During Lateral}} and {{Longitudinal
  Assistance}}} {A {{Reference Model}} for {{Driver Attention}} in
  {{Automation}}: {{Glance Behavior Changes During Lateral}} and {{Longitudinal
  Assistance}}}.{\BBCQ}
\newblock
\APACjournalVolNumPages{IEEE Transactions on Intelligent Transportation
  Systems}{20}{8}{2999--3009}.
\newblock
\begin{APACrefDOI} \doi{10.1109/TITS.2018.2870909} \end{APACrefDOI}
\PrintBackRefs{\CurrentBib}

\bibitem [\protect \citeauthoryear {%
Morgenstern%
, Schott%
\BCBL {}\ \BBA {} Krems%
}{%
Morgenstern%
\ \protect \BOthers {.}}{%
{\protect \APACyear {2020}}%
}]{%
morgenstern.2020}
\APACinsertmetastar {%
morgenstern.2020}%
\begin{APACrefauthors}%
Morgenstern, T.%
, Schott, L.%
\BCBL {}\ \BBA {} Krems, J\BPBI F.%
\end{APACrefauthors}%
\unskip\
\newblock
\APACrefYearMonthDay{2020}{{\APACmonth{08}}}{}.
\newblock
{\BBOQ}\APACrefatitle {Do Drivers Reduce Their Speed When Texting on Highways?
  {{A}} Replication Study Using {{European}} Naturalistic Driving Data} {Do
  drivers reduce their speed when texting on highways? {{A}} replication study
  using {{European}} naturalistic driving data}.{\BBCQ}
\newblock
\APACjournalVolNumPages{Safety Science}{128}{}{104740}.
\newblock
\begin{APACrefDOI} \doi{10.1016/j.ssci.2020.104740} \end{APACrefDOI}
\PrintBackRefs{\CurrentBib}

\bibitem [\protect \citeauthoryear {%
Nakagawa%
\ \BBA {} Schielzeth%
}{%
Nakagawa%
\ \BBA {} Schielzeth%
}{%
{\protect \APACyear {2013}}%
}]{%
Nakagawa.2013}
\APACinsertmetastar {%
Nakagawa.2013}%
\begin{APACrefauthors}%
Nakagawa, S.%
\BCBT {}\ \BBA {} Schielzeth, H.%
\end{APACrefauthors}%
\unskip\
\newblock
\APACrefYearMonthDay{2013}{{\APACmonth{02}}}{}.
\newblock
{\BBOQ}\APACrefatitle {A General and Simple Method for Obtaining {{{\emph{R}}}}
  {\textsuperscript{2}} from Generalized Linear Mixed-Effects Models} {A
  general and simple method for obtaining {{{\emph{R}}}} {\textsuperscript{2}}
  from generalized linear mixed-effects models}.{\BBCQ}
\newblock
\APACjournalVolNumPages{Methods in Ecology and Evolution}{4}{2}{133--142}.
\newblock
\begin{APACrefDOI} \doi{10.1111/j.2041-210x.2012.00261.x} \end{APACrefDOI}
\PrintBackRefs{\CurrentBib}

\bibitem [\protect \citeauthoryear {%
{National Center for Statistics {and} Analysis}%
}{%
{National Center for Statistics {and} Analysis}%
}{%
{\protect \APACyear {2014}}%
}]{%
NHTSA.2014}
\APACinsertmetastar {%
NHTSA.2014}%
\begin{APACrefauthors}%
{National Center for Statistics {and} Analysis}.%
\end{APACrefauthors}%
\unskip\
\newblock
\APACrefYearMonthDay{2014}{}{}.
\newblock
\APACrefbtitle {Visual-{{Manual NHTSA Driver Distraction Guidelines}} for
  {{In-Vehicle Electronic Devices}}} {Visual-{{Manual NHTSA Driver Distraction
  Guidelines}} for {{In-Vehicle Electronic Devices}}}\ \APACbVolEdTR {}{Notice\
  \BNUM\ NHTSA- 2014-0088}.
\newblock
\APACaddressInstitution{}{{National Highway Traffic Safety Administration}}.
\PrintBackRefs{\CurrentBib}

\bibitem [\protect \citeauthoryear {%
{National Center for Statistics and Analysis}%
}{%
{National Center for Statistics and Analysis}%
}{%
{\protect \APACyear {2021}}%
}]{%
NHTSA.2021}
\APACinsertmetastar {%
NHTSA.2021}%
\begin{APACrefauthors}%
{National Center for Statistics and Analysis}.%
\end{APACrefauthors}%
\unskip\
\newblock
\APACrefYearMonthDay{2021}{}{}.
\newblock
\APACrefbtitle {Distracted Driving 2019} {Distracted driving 2019}\
  \APACbVolEdTR{}{\BTR{}\ \BNUM\ Report No. DOT HS 811 299}.
\newblock
\APACaddressInstitution{}{{National Highway Traffic Safety Administration}}.
\PrintBackRefs{\CurrentBib}

\bibitem [\protect \citeauthoryear {%
Naujoks%
, Purucker%
\BCBL {}\ \BBA {} Neukum%
}{%
Naujoks%
\ \protect \BOthers {.}}{%
{\protect \APACyear {2016}}%
}]{%
naujoks.2016}
\APACinsertmetastar {%
naujoks.2016}%
\begin{APACrefauthors}%
Naujoks, F.%
, Purucker, C.%
\BCBL {}\ \BBA {} Neukum, A.%
\end{APACrefauthors}%
\unskip\
\newblock
\APACrefYearMonthDay{2016}{{\APACmonth{04}}}{}.
\newblock
{\BBOQ}\APACrefatitle {Secondary Task Engagement and Vehicle Automation
  \textendash{} {{Comparing}} the Effects of Different Automation Levels in an
  on-Road Experiment} {Secondary task engagement and vehicle automation
  \textendash{} {{Comparing}} the effects of different automation levels in an
  on-road experiment}.{\BBCQ}
\newblock
\APACjournalVolNumPages{Transportation Research Part F: Traffic Psychology and
  Behaviour}{38}{}{67--82}.
\newblock
\begin{APACrefDOI} \doi{10.1016/j.trf.2016.01.011} \end{APACrefDOI}
\PrintBackRefs{\CurrentBib}

\bibitem [\protect \citeauthoryear {%
Noble%
, Miles%
, Perez%
, Guo%
\BCBL {}\ \BBA {} Klauer%
}{%
Noble%
\ \protect \BOthers {.}}{%
{\protect \APACyear {2021}}%
}]{%
noble.2021}
\APACinsertmetastar {%
noble.2021}%
\begin{APACrefauthors}%
Noble, A\BPBI M.%
, Miles, M.%
, Perez, M\BPBI A.%
, Guo, F.%
\BCBL {}\ \BBA {} Klauer, S\BPBI G.%
\end{APACrefauthors}%
\unskip\
\newblock
\APACrefYearMonthDay{2021}{{\APACmonth{03}}}{}.
\newblock
{\BBOQ}\APACrefatitle {Evaluating Driver Eye Glance Behavior and Secondary Task
  Engagement While Using Driving Automation Systems} {Evaluating driver eye
  glance behavior and secondary task engagement while using driving automation
  systems}.{\BBCQ}
\newblock
\APACjournalVolNumPages{Accident Analysis \& Prevention}{151}{}{105959}.
\newblock
\begin{APACrefDOI} \doi{10.1016/j.aap.2020.105959} \end{APACrefDOI}
\PrintBackRefs{\CurrentBib}

\bibitem [\protect \citeauthoryear {%
{On-Road Automated Driving (ORAD) committee}%
}{%
{On-Road Automated Driving (ORAD) committee}%
}{%
{\protect \APACyear {2021}}%
}]{%
SAE.2021}
\APACinsertmetastar {%
SAE.2021}%
\begin{APACrefauthors}%
{On-Road Automated Driving (ORAD) committee}.%
\end{APACrefauthors}%
\unskip\
\newblock
\APACrefYearMonthDay{2021}{}{}.
\newblock
\APACrefbtitle {Taxonomy and {{Definitions}} for {{Terms Related}} to {{Driving
  Automation Systems}} for {{On-Road Motor Vehicles}}} {Taxonomy and
  {{Definitions}} for {{Terms Related}} to {{Driving Automation Systems}} for
  {{On-Road Motor Vehicles}}}\ \APACbVolEdTR{}{\BTR{}}.
\newblock
\APACaddressInstitution{}{{SAE International}}.
\newblock
\begin{APACrefDOI} \doi{10.4271/J3016_202104} \end{APACrefDOI}
\PrintBackRefs{\CurrentBib}

\bibitem [\protect \citeauthoryear {%
{Onate-Vega}%
, {Oviedo-Trespalacios}%
\BCBL {}\ \BBA {} King%
}{%
{Onate-Vega}%
\ \protect \BOthers {.}}{%
{\protect \APACyear {2020}}%
}]{%
onate-vega.2020}
\APACinsertmetastar {%
onate-vega.2020}%
\begin{APACrefauthors}%
{Onate-Vega}, D.%
, {Oviedo-Trespalacios}, O.%
\BCBL {}\ \BBA {} King, M\BPBI J.%
\end{APACrefauthors}%
\unskip\
\newblock
\APACrefYearMonthDay{2020}{{\APACmonth{05}}}{}.
\newblock
{\BBOQ}\APACrefatitle {How Drivers Adapt Their Behaviour to Changes in Task
  Complexity: {{The}} Role of Secondary Task Demands and Road Environment
  Factors} {How drivers adapt their behaviour to changes in task complexity:
  {{The}} role of secondary task demands and road environment factors}.{\BBCQ}
\newblock
\APACjournalVolNumPages{Transportation Research Part F: Traffic Psychology and
  Behaviour}{71}{}{145--156}.
\newblock
\begin{APACrefDOI} \doi{10.1016/j.trf.2020.03.015} \end{APACrefDOI}
\PrintBackRefs{\CurrentBib}

\bibitem [\protect \citeauthoryear {%
{Oviedo-Trespalacios}%
, Haque%
, King%
\BCBL {}\ \BBA {} Demmel%
}{%
{Oviedo-Trespalacios}%
, Haque%
, King%
\BCBL {}\ \BBA {} Demmel%
}{%
{\protect \APACyear {2018}}%
}]{%
oviedo-trespalacios.2018}
\APACinsertmetastar {%
oviedo-trespalacios.2018}%
\begin{APACrefauthors}%
{Oviedo-Trespalacios}, O.%
, Haque, M\BPBI M.%
, King, M.%
\BCBL {}\ \BBA {} Demmel, S.%
\end{APACrefauthors}%
\unskip\
\newblock
\APACrefYearMonthDay{2018}{{\APACmonth{09}}}{}.
\newblock
{\BBOQ}\APACrefatitle {Driving Behaviour While Self-Regulating Mobile Phone
  Interactions: {{A}} Human-Machine System Approach} {Driving behaviour while
  self-regulating mobile phone interactions: {{A}} human-machine system
  approach}.{\BBCQ}
\newblock
\APACjournalVolNumPages{Accident Analysis \& Prevention}{118}{}{253--262}.
\newblock
\begin{APACrefDOI} \doi{10.1016/j.aap.2018.03.020} \end{APACrefDOI}
\PrintBackRefs{\CurrentBib}

\bibitem [\protect \citeauthoryear {%
{Oviedo-Trespalacios}%
, Haque%
, King%
\BCBL {}\ \BBA {} Washington%
}{%
{Oviedo-Trespalacios}%
, Haque%
, King%
\BCBL {}\ \BBA {} Washington%
}{%
{\protect \APACyear {2018}}%
}]{%
oviedo-trespalacios.2018a}
\APACinsertmetastar {%
oviedo-trespalacios.2018a}%
\begin{APACrefauthors}%
{Oviedo-Trespalacios}, O.%
, Haque, M\BPBI M.%
, King, M.%
\BCBL {}\ \BBA {} Washington, S.%
\end{APACrefauthors}%
\unskip\
\newblock
\APACrefYearMonthDay{2018}{{\APACmonth{10}}}{}.
\newblock
{\BBOQ}\APACrefatitle {Should {{I Text}} or {{Call Here}}? {{A Situation-Based
  Analysis}} of {{Drivers}}' {{Perceived Likelihood}} of {{Engaging}} in
  {{Mobile Phone Multitasking}}: {{Mobile Phone Multitasking Engagement}}}
  {Should {{I Text}} or {{Call Here}}? {{A Situation-Based Analysis}} of
  {{Drivers}}' {{Perceived Likelihood}} of {{Engaging}} in {{Mobile Phone
  Multitasking}}: {{Mobile Phone Multitasking Engagement}}}.{\BBCQ}
\newblock
\APACjournalVolNumPages{Risk Analysis}{38}{10}{2144--2160}.
\newblock
\begin{APACrefDOI} \doi{10.1111/risa.13119} \end{APACrefDOI}
\PrintBackRefs{\CurrentBib}

\bibitem [\protect \citeauthoryear {%
{Oviedo-Trespalacios}%
, Haque%
, King%
\BCBL {}\ \BBA {} Washington%
}{%
{Oviedo-Trespalacios}%
\ \protect \BOthers {.}}{%
{\protect \APACyear {2019}}%
}]{%
oviedo-trespalacios.2019}
\APACinsertmetastar {%
oviedo-trespalacios.2019}%
\begin{APACrefauthors}%
{Oviedo-Trespalacios}, O.%
, Haque, M\BPBI M.%
, King, M.%
\BCBL {}\ \BBA {} Washington, S.%
\end{APACrefauthors}%
\unskip\
\newblock
\APACrefYearMonthDay{2019}{{\APACmonth{01}}}{}.
\newblock
{\BBOQ}\APACrefatitle {``{{Mate}}! {{I}}'m Running 10 Min Late'': {{An}}
  Investigation into the Self-Regulation of Mobile Phone Tasks While Driving}
  {``{{Mate}}! {{I}}'m running 10 min late'': {{An}} investigation into the
  self-regulation of mobile phone tasks while driving}.{\BBCQ}
\newblock
\APACjournalVolNumPages{Accident Analysis \& Prevention}{122}{}{134--142}.
\newblock
\begin{APACrefDOI} \doi{10.1016/j.aap.2018.09.020} \end{APACrefDOI}
\PrintBackRefs{\CurrentBib}

\bibitem [\protect \citeauthoryear {%
{R Core Team}%
}{%
{R Core Team}%
}{%
{\protect \APACyear {2022}}%
}]{%
rcoreteam.2022}
\APACinsertmetastar {%
rcoreteam.2022}%
\begin{APACrefauthors}%
{R Core Team}.%
\end{APACrefauthors}%
\unskip\
\newblock
\APACrefYearMonthDay{2022}{}{}.
\newblock
\APACrefbtitle {R: {{A}} Language and Environment for Statistical Computing}
  {R: {{A}} language and environment for statistical computing}\ [Manual].
\newblock
\APACaddressPublisher{{Vienna, Austria}}{}.
\PrintBackRefs{\CurrentBib}

\bibitem [\protect \citeauthoryear {%
Regan%
, Lee%
\BCBL {}\ \BBA {} Young%
}{%
Regan%
\ \protect \BOthers {.}}{%
{\protect \APACyear {2009}}%
}]{%
regan.2009}
\APACinsertmetastar {%
regan.2009}%
\begin{APACrefauthors}%
Regan, M\BPBI A.%
, Lee, J\BPBI D.%
\BCBL {}\ \BBA {} Young, K\BPBI L.%
\end{APACrefauthors}%
\ (\BEDS).
\unskip\
\newblock
\APACrefYear{2009}.
\newblock
\APACrefbtitle {Driver Distraction: Theory, Effects, and Mitigation} {Driver
  distraction: Theory, effects, and mitigation}.
\newblock
\APACaddressPublisher{{Boca Raton}}{{CRC Press/Taylor \& Francis Group}}.
\PrintBackRefs{\CurrentBib}

\bibitem [\protect \citeauthoryear {%
Regan%
\ \BBA {} {Oviedo-Trespalacios}%
}{%
Regan%
\ \BBA {} {Oviedo-Trespalacios}%
}{%
{\protect \APACyear {2022}}%
}]{%
regan.2022}
\APACinsertmetastar {%
regan.2022}%
\begin{APACrefauthors}%
Regan, M\BPBI A.%
\BCBT {}\ \BBA {} {Oviedo-Trespalacios}, O.%
\end{APACrefauthors}%
\unskip\
\newblock
\APACrefYearMonthDay{2022}{}{}.
\newblock
{\BBOQ}\APACrefatitle {Driver {{Distraction}}: {{Mechanisms}}, {{Evidence}},
  {{Prevention}}, and {{Mitigation}}} {Driver {{Distraction}}: {{Mechanisms}},
  {{Evidence}}, {{Prevention}}, and {{Mitigation}}}.{\BBCQ}
\newblock
\BIn{} K.~Edvardsson~Bj{\"o}rnberg, M\BHBI {\AA}.~Belin, S\BPBI O.~Hansson\BCBL
  {}\ \BBA {} C.~Tingvall\ (\BEDS), \APACrefbtitle {The {{Vision Zero
  Handbook}}} {The {{Vision Zero Handbook}}}\ (\BPGS\ 1--62).
\newblock
\APACaddressPublisher{{Cham}}{{Springer International Publishing}}.
\newblock
\begin{APACrefDOI} \doi{10.1007/978-3-030-23176-7_38-1} \end{APACrefDOI}
\PrintBackRefs{\CurrentBib}

\bibitem [\protect \citeauthoryear {%
Risteska%
, Kanaan%
, Donmez%
\BCBL {}\ \BBA {} Winnie~Chen%
}{%
Risteska%
\ \protect \BOthers {.}}{%
{\protect \APACyear {2021}}%
}]{%
risteska.2021}
\APACinsertmetastar {%
risteska.2021}%
\begin{APACrefauthors}%
Risteska, M.%
, Kanaan, D.%
, Donmez, B.%
\BCBL {}\ \BBA {} Winnie~Chen, H\BHBI Y.%
\end{APACrefauthors}%
\unskip\
\newblock
\APACrefYearMonthDay{2021}{{\APACmonth{04}}}{}.
\newblock
{\BBOQ}\APACrefatitle {The Effect of Driving Demands on Distraction Engagement
  and Glance Behaviors: {{Results}} from Naturalistic Data} {The effect of
  driving demands on distraction engagement and glance behaviors: {{Results}}
  from naturalistic data}.{\BBCQ}
\newblock
\APACjournalVolNumPages{Safety Science}{136}{}{105123}.
\newblock
\begin{APACrefDOI} \doi{10.1016/j.ssci.2020.105123} \end{APACrefDOI}
\PrintBackRefs{\CurrentBib}

\bibitem [\protect \citeauthoryear {%
{Rudin-Brown}%
}{%
{Rudin-Brown}%
}{%
{\protect \APACyear {2013}}%
}]{%
rudin-brown.2013a}
\APACinsertmetastar {%
rudin-brown.2013a}%
\begin{APACrefauthors}%
{Rudin-Brown}, C.%
\end{APACrefauthors}%
\ (\BED).
\unskip\
\newblock
\APACrefYear{2013}.
\newblock
\APACrefbtitle {Behavioural Adaptation and Road Safety: Theory, Evidence, and
  Action} {Behavioural adaptation and road safety: Theory, evidence, and
  action}.
\newblock
\APACaddressPublisher{{Boca Raton}}{{CRC Press, Taylor \& Francis Group}}.
\PrintBackRefs{\CurrentBib}

\bibitem [\protect \citeauthoryear {%
Schneidereit%
, Petzoldt%
, Keinath%
\BCBL {}\ \BBA {} Krems%
}{%
Schneidereit%
\ \protect \BOthers {.}}{%
{\protect \APACyear {2017}}%
}]{%
schneidereit.2017}
\APACinsertmetastar {%
schneidereit.2017}%
\begin{APACrefauthors}%
Schneidereit, T.%
, Petzoldt, T.%
, Keinath, A.%
\BCBL {}\ \BBA {} Krems, J\BPBI F.%
\end{APACrefauthors}%
\unskip\
\newblock
\APACrefYearMonthDay{2017}{{\APACmonth{09}}}{}.
\newblock
{\BBOQ}\APACrefatitle {Using {{SHRP}} 2 Naturalistic Driving Data to Assess
  Drivers' Speed Choice While Being Engaged in Different Secondary Tasks}
  {Using {{SHRP}} 2 naturalistic driving data to assess drivers' speed choice
  while being engaged in different secondary tasks}.{\BBCQ}
\newblock
\APACjournalVolNumPages{Journal of Safety Research}{62}{}{33--42}.
\newblock
\begin{APACrefDOI} \doi{10.1016/j.jsr.2017.04.004} \end{APACrefDOI}
\PrintBackRefs{\CurrentBib}

\bibitem [\protect \citeauthoryear {%
Sch{\"o}mig%
\ \BBA {} Metz%
}{%
Sch{\"o}mig%
\ \BBA {} Metz%
}{%
{\protect \APACyear {2013}}%
}]{%
schomig.2013}
\APACinsertmetastar {%
schomig.2013}%
\begin{APACrefauthors}%
Sch{\"o}mig, N.%
\BCBT {}\ \BBA {} Metz, B.%
\end{APACrefauthors}%
\unskip\
\newblock
\APACrefYearMonthDay{2013}{{\APACmonth{07}}}{}.
\newblock
{\BBOQ}\APACrefatitle {Three Levels of Situation Awareness in Driving with
  Secondary Tasks} {Three levels of situation awareness in driving with
  secondary tasks}.{\BBCQ}
\newblock
\APACjournalVolNumPages{Safety Science}{56}{}{44--51}.
\newblock
\begin{APACrefDOI} \doi{10.1016/j.ssci.2012.05.029} \end{APACrefDOI}
\PrintBackRefs{\CurrentBib}

\bibitem [\protect \citeauthoryear {%
Starkey%
\ \BBA {} Charlton%
}{%
Starkey%
\ \BBA {} Charlton%
}{%
{\protect \APACyear {2020}}%
}]{%
starkey.2020}
\APACinsertmetastar {%
starkey.2020}%
\begin{APACrefauthors}%
Starkey, N\BPBI J.%
\BCBT {}\ \BBA {} Charlton, S\BPBI G.%
\end{APACrefauthors}%
\unskip\
\newblock
\APACrefYearMonthDay{2020}{{\APACmonth{08}}}{}.
\newblock
{\BBOQ}\APACrefatitle {Drivers {{Use}} of {{In-Vehicle Information Systems}}
  and {{Perceptions}} of {{Their Effects}} on {{Driving}}} {Drivers {{Use}} of
  {{In-Vehicle Information Systems}} and {{Perceptions}} of {{Their Effects}}
  on {{Driving}}}.{\BBCQ}
\newblock
\APACjournalVolNumPages{Frontiers in Sustainable Cities}{2}{}{39}.
\newblock
\begin{APACrefDOI} \doi{10.3389/frsc.2020.00039} \end{APACrefDOI}
\PrintBackRefs{\CurrentBib}

\bibitem [\protect \citeauthoryear {%
Tivesten%
\ \BBA {} Dozza%
}{%
Tivesten%
\ \BBA {} Dozza%
}{%
{\protect \APACyear {2014}}%
}]{%
tivesten.2014}
\APACinsertmetastar {%
tivesten.2014}%
\begin{APACrefauthors}%
Tivesten, E.%
\BCBT {}\ \BBA {} Dozza, M.%
\end{APACrefauthors}%
\unskip\
\newblock
\APACrefYearMonthDay{2014}{{\APACmonth{09}}}{}.
\newblock
{\BBOQ}\APACrefatitle {Driving Context and Visual-Manual Phone Tasks Influence
  Glance Behavior in Naturalistic Driving} {Driving context and visual-manual
  phone tasks influence glance behavior in naturalistic driving}.{\BBCQ}
\newblock
\APACjournalVolNumPages{Transportation Research Part F: Traffic Psychology and
  Behaviour}{26}{}{258--272}.
\newblock
\begin{APACrefDOI} \doi{10.1016/j.trf.2014.08.004} \end{APACrefDOI}
\PrintBackRefs{\CurrentBib}

\bibitem [\protect \citeauthoryear {%
Tivesten%
\ \BBA {} Dozza%
}{%
Tivesten%
\ \BBA {} Dozza%
}{%
{\protect \APACyear {2015}}%
}]{%
tivesten.2015}
\APACinsertmetastar {%
tivesten.2015}%
\begin{APACrefauthors}%
Tivesten, E.%
\BCBT {}\ \BBA {} Dozza, M.%
\end{APACrefauthors}%
\unskip\
\newblock
\APACrefYearMonthDay{2015}{{\APACmonth{06}}}{}.
\newblock
{\BBOQ}\APACrefatitle {Driving Context Influences Drivers' Decision to Engage
  in Visual\textendash Manual Phone Tasks: {{Evidence}} from a Naturalistic
  Driving Study} {Driving context influences drivers' decision to engage in
  visual\textendash manual phone tasks: {{Evidence}} from a naturalistic
  driving study}.{\BBCQ}
\newblock
\APACjournalVolNumPages{Journal of Safety Research}{53}{}{87--96}.
\newblock
\begin{APACrefDOI} \doi{10.1016/j.jsr.2015.03.010} \end{APACrefDOI}
\PrintBackRefs{\CurrentBib}

\bibitem [\protect \citeauthoryear {%
Tukey%
}{%
Tukey%
}{%
{\protect \APACyear {1949}}%
}]{%
tukey.1949}
\APACinsertmetastar {%
tukey.1949}%
\begin{APACrefauthors}%
Tukey, J\BPBI W.%
\end{APACrefauthors}%
\unskip\
\newblock
\APACrefYearMonthDay{1949}{{\APACmonth{06}}}{}.
\newblock
{\BBOQ}\APACrefatitle {Comparing {{Individual Means}} in the {{Analysis}} of
  {{Variance}}} {Comparing {{Individual Means}} in the {{Analysis}} of
  {{Variance}}}.{\BBCQ}
\newblock
\APACjournalVolNumPages{Biometrics}{5}{2}{99}.
\newblock
\begin{APACrefDOI} \doi{10.2307/3001913} \end{APACrefDOI}
\PrintBackRefs{\CurrentBib}

\bibitem [\protect \citeauthoryear {%
Wikman%
, Nieminen%
\BCBL {}\ \BBA {} Summala%
}{%
Wikman%
\ \protect \BOthers {.}}{%
{\protect \APACyear {1998}}%
}]{%
wikman.1998}
\APACinsertmetastar {%
wikman.1998}%
\begin{APACrefauthors}%
Wikman, A\BHBI S.%
, Nieminen, T.%
\BCBL {}\ \BBA {} Summala, H.%
\end{APACrefauthors}%
\unskip\
\newblock
\APACrefYearMonthDay{1998}{{\APACmonth{03}}}{}.
\newblock
{\BBOQ}\APACrefatitle {Driving Experience and Time-Sharing during in-Car Tasks
  on Roads of Different Width} {Driving experience and time-sharing during
  in-car tasks on roads of different width}.{\BBCQ}
\newblock
\APACjournalVolNumPages{Ergonomics}{41}{3}{358--372}.
\newblock
\begin{APACrefDOI} \doi{10.1080/001401398187080} \end{APACrefDOI}
\PrintBackRefs{\CurrentBib}

\bibitem [\protect \citeauthoryear {%
Yang%
, Kuo%
\BCBL {}\ \BBA {} Lenn{\'e}%
}{%
Yang%
\ \protect \BOthers {.}}{%
{\protect \APACyear {2021}}%
}]{%
yang.2021}
\APACinsertmetastar {%
yang.2021}%
\begin{APACrefauthors}%
Yang, S.%
, Kuo, J.%
\BCBL {}\ \BBA {} Lenn{\'e}, M\BPBI G.%
\end{APACrefauthors}%
\unskip\
\newblock
\APACrefYearMonthDay{2021}{{\APACmonth{12}}}{}.
\newblock
{\BBOQ}\APACrefatitle {Effects of {{Distraction}} in {{On-Road Level}} 2
  {{Automated Driving}}: {{Impacts}} on {{Glance Behavior}} and {{Takeover
  Performance}}} {Effects of {{Distraction}} in {{On-Road Level}} 2 {{Automated
  Driving}}: {{Impacts}} on {{Glance Behavior}} and {{Takeover
  Performance}}}.{\BBCQ}
\newblock
\APACjournalVolNumPages{Human Factors: The Journal of the Human Factors and
  Ergonomics Society}{63}{8}{1485--1497}.
\newblock
\begin{APACrefDOI} \doi{10.1177/0018720820936793} \end{APACrefDOI}
\PrintBackRefs{\CurrentBib}

\bibitem [\protect \citeauthoryear {%
Young%
\ \BBA {} Lenn{\'e}%
}{%
Young%
\ \BBA {} Lenn{\'e}%
}{%
{\protect \APACyear {2010}}%
}]{%
young.2010}
\APACinsertmetastar {%
young.2010}%
\begin{APACrefauthors}%
Young, K\BPBI L.%
\BCBT {}\ \BBA {} Lenn{\'e}, M\BPBI G.%
\end{APACrefauthors}%
\unskip\
\newblock
\APACrefYearMonthDay{2010}{{\APACmonth{03}}}{}.
\newblock
{\BBOQ}\APACrefatitle {Driver Engagement in Distracting Activities and the
  Strategies Used to Minimise Risk} {Driver engagement in distracting
  activities and the strategies used to minimise risk}.{\BBCQ}
\newblock
\APACjournalVolNumPages{Safety Science}{48}{3}{326--332}.
\newblock
\begin{APACrefDOI} \doi{10.1016/j.ssci.2009.10.008} \end{APACrefDOI}
\PrintBackRefs{\CurrentBib}

\end{thebibliography}

\section*{About the authors}

\textbf{Patrick Ebel} is a PhD student at the Department of Software and Systems Engineering at the University of Cologne. He received his master's degree in automotive systems from the TU Berlin in 2019. His research focuses on the analysis of large naturalistic driving data and computational models for driver interactions with IVIS. \\ 

\noindent
\textbf{Christoph Lingenfelder} received his PhD from the University of Kaiserslautern in 1990. He spent many years researching and developing data mining methods at IBM Research, and is now Lead Artificial Intelligence at MBition, a Mercedes-Benz subsidiary working on the next-generation infotainment systems.\\

\noindent
\textbf{Andreas Vogelsang} is Professor of Software and Systems Engineering at the University of Cologne. Before, he was Junior Professor for Automotive Software Engineering at TU Berlin and Head of Software Engineering at the Daimler Center for Automotive IT Innovations. His research focuses on requirements engineering and data-driven systems engineering.\\

\section{Appendices}

\appendix
\section{Appendix}\label{ch:Appendix}

\subsection{Dataset Summary Statistics}\label{ch:AppendixSummaryStats}

\begin{table}[!htbp] \centering 
  \caption{Summary statistics over all 31,378 interaction sequences.} 
  \footnotesize
  \label{tab:DatasetStatistics} 
\begin{tabular}{@{\extracolsep{5pt}}lrrrrrrr} 
\\[-1.8ex]\hline 
\hline \\[-1.8ex] 
Statistic & \multicolumn{1}{c}{Mean} & \multicolumn{1}{c}{St. Dev.} & \multicolumn{1}{c}{Min} & \multicolumn{1}{c}{Pctl(25)} & \multicolumn{1}{c}{Median} & \multicolumn{1}{c}{Pctl(75)} & \multicolumn{1}{c}{Max} \\ 
\hline \\[-1.8ex] 
Number of interactions & 6.515 & 5.530 & 1 & 3 & 5 & 8 & 41 \\ 
Number of tap gestures & 4.988 & 4.913 & 0 & 2 & 4 & 6 & 40 \\ 
Number of drag gestures & 0.789 & 1.826 & 0 & 0 & 0 & 1 & 32 \\ 
Number of multitouch gestures & 0.721 & 1.992 & 0 & 0 & 0 & 0 & 29 \\ 
Mean glance duration center stack in s& 1.72 & 1.23 & 0.12 & 1.03 & 1.40 & 2.00 & 34.08 \\ 
Number of glances per sequence & 6.068 & 5.106 & 1 & 3 & 5 & 8 & 59 \\ 
Number of long glances & 1.322 & 1.875 & 0 & 0 & 1 & 2 & 22 \\ 
Total glance duration in s & 9.70 & 9.26 & 0.12 & 3.96 & 7.08 & 12.16 & 262.42 \\ 
Average speed in km/h & 77.57 & 35.95 & 0.37 & 48.33 & 77.71 & 104.61 & 242.26 \\ 
Number of Keyboard interactions & 0.330 & 2.018 & 0 & 0 & 0 & 0 & 37 \\ 
Number of Tab interactions & 0.314 & 1.081 & 0 & 0 & 0 & 0 & 35 \\ 
Number of List interactions & 0.852 & 2.097 & 0 & 0 & 0 & 1 & 41 \\ 
Number of Map interactions & 1.394 & 3.581 & 0 & 0 & 0 & 1 & 40 \\ 
Number of ControlBar interactions & 0.020 & 0.165 & 0 & 0 & 0 & 0 & 4 \\ 
Number of Button interactions & 0.862 & 1.912 & 0 & 0 & 0 & 1 & 36 \\ 
Number of Homebar interactions & 1.126 & 2.341 & 0 & 0 & 0 & 1 & 36 \\ 
Number of ClickGuard interactions & 0.063 & 0.343 & 0 & 0 & 0 & 0 & 14 \\ 
Number of CoverFlow interactions & 0.107 & 0.806 & 0 & 0 & 0 & 0 & 27 \\ 
Number of PopUp interactions & 0.019 & 0.176 & 0 & 0 & 0 & 0 & 9 \\ 
Number of AppIcon interactions & 0.286 & 0.644 & 0 & 0 & 0 & 0 & 14 \\ 
Number of Slider interactions & 0.062 & 0.481 & 0 & 0 & 0 & 0 & 22 \\ 
Number of Other interactions & 0.833 & 1.626 & 0 & 0 & 0 & 1 & 37 \\ 
Number of Unknown interactions & 0.030 & 0.347 & 0 & 0 & 0 & 0 & 23 \\ 
Number of RemoteUI interactions & 0.215 & 1.413 & 0 & 0 & 0 & 0 & 40 \\ 
\hline \\[-1.8ex] 
\end{tabular} 
\end{table} 

\newpage
\subsection{Models}\label{ch:AppendixModels}

\begin{sidewaystable}[!htbp] \centering 

\caption{Mixed-effects models for the interaction probability with Keyboard, CoverFlow, Slider, RemoteUI, ControlBar, Other, and Unknown UI elements. Note: In contrast to the models presented in the paper we did not include the car type as random effect since it led to a singularity warning. This warning is often associated with an overfitted model as the random effect structure might be too complex to be supported by the data. This in turn might be due to the small amount of interaction with these UI elements.}
\label{} 
\begin{tabular}{@{\extracolsep{5pt}}lrrrrrr} 
\\[-1.8ex]\hline 
\hline \\[-1.8ex] 
 & \multicolumn{6}{c}{\textit{Dependent variable:}} \\ 
\cline{2-7} 
\\[-1.8ex] & Keyboard & CoverFlow & Slider & RemoteUI & ControlBar & Other \\ 
\\[-1.8ex] &  &  & \textit{conv. error} &  &  & \\ 
\hline \\[-1.8ex] 
Intercept & $-$8.46$^{***}$ (0.19) & $-$9.35$^{***}$ (0.22) & $-$8.88$^{***}$ (1) & $-$10.33$^{***}$ (0.23) & $-$10.37$^{***}$ (0.30) & $-$0.40$^{***}$ (0.03) \\ 
  ACC & 0.07 (0.26) & 0.43 (0.30) & 0.02$^{***}$ (1) & $-$0.01 (0.30) & 0.31 (0.34) & 0.06 (0.07) \\ 
  ACC+LKA & 0.21 (0.14) & 0.39$^{*}$ (0.16) & 0.18 (0.12) & $-$0.31$^{*}$ (0.15) & 0.14 (0.22) & $-$0.17$^{***}$ (0.03) \\ 
  50-100 & $-$0.19 (0.12) & $-$0.13 (0.14) & 0.24$^{***}$ (1) & $-$0.07 (0.15) & 0.32 (0.18) & 0.07$^{*}$ (0.03) \\ 
  100+ & $-$0.32$^{*}$ (0.14) & $-$0.31 (0.16) & 0.23$^{*}$ (0.11) & 0.05 (0.17) & 0.37 (0.22) & 0.09$^{*}$ (0.04) \\ 
  curved & $-$0.15 (0.13) & $-$0.38$^{*}$ (0.16) & $-$0.33$^{*}$ (0.15) & $-$0.14 (0.15) & $-$0.36 (0.20) & $-$0.06 (0.04) \\ 
 \hline \\[-1.8ex] 
Akaike Inf. Crit. & 8,886.07 & 6,982.70 & 7,282.68 & 6,270.20 & 3,965.26 & 41,098.95 \\ 
Bayesian Inf. Crit. & 8,944.55 & 7,041.17 & 7,341.16 & 6,328.68 & 4,023.74 & 41,157.43 \\ 
\hline 
\hline \\[-1.8ex] 
\textit{Note:}  & \multicolumn{6}{r}{conv. error: Model failed to converge, $^{*}$p$<$0.05; $^{**}$p$<$0.01; $^{***}$p$<$0.001} \\ 
\end{tabular}
\end{sidewaystable}

\begin{table}[!htbp] \centering 
  \caption{Mixed effects models for the mean glance duration according to speed and road curvature.} 
  \small
  \label{tab:SpeedCurvatureOnly} 
\begin{tabular}{@{\extracolsep{5pt}}lrr} 
\\[-1.8ex]\hline 
\hline \\[-1.8ex] 
 & \multicolumn{2}{c}{\textit{Dependent variable:}} \\ 
\cline{2-3} 
\\[-1.8ex] & \multicolumn{2}{c}{Mean Glance Duration} \\ 
\\[-1.8ex] & Model 5 & Model 6 \\ 
\hline \\[-1.8ex] 
 Constant & 7.28$^{***}$ (0.02) & 7.27$^{***}$ (0.02) \\ 
  50-100 & $-$0.06$^{***}$ (0.01) &  \\ 
  100+ & $-$0.07$^{***}$ (0.01) &  \\ 
  curved &  & $-$0.14$^{***}$ (0.01) \\ 
 \hline \\[-1.8ex] 
Akaike Inf. Crit. & 44,429.86 & 44,178.02 \\ 
Bayesian Inf. Crit. & 44,479.98 & 44,219.78 \\ 
\hline 
\hline \\[-1.8ex] 
\textit{Note:}  & \multicolumn{2}{r}{$^{*}$p$<$0.05; $^{**}$p$<$0.01; $^{***}$p$<$0.001} \\ 
\end{tabular} 
\end{table} 

\begin{figure*}
	\centering
	\includegraphics[width=0.6\linewidth]{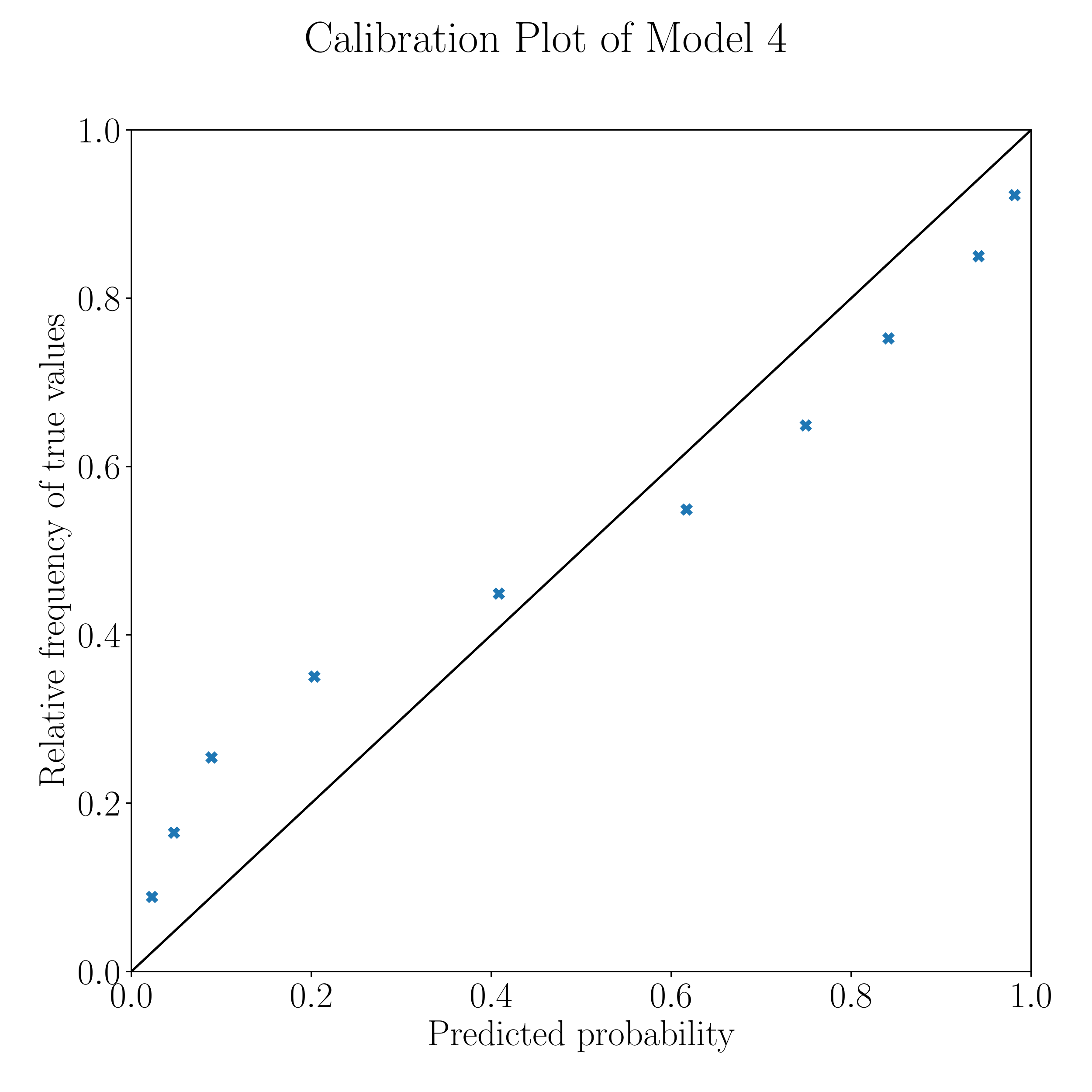} 
	\caption{Calibration plot of Model 4 (generalized linear mixed-effects model). Marginal $R^2 = 0.099$, Conditional $R^2 = 	0.347$} 
	\label{fig:CalibrationPlot} 
\end{figure*}

\begin{figure*}[!htbp]
    \centering
    \includegraphics[width = 0.6\linewidth]{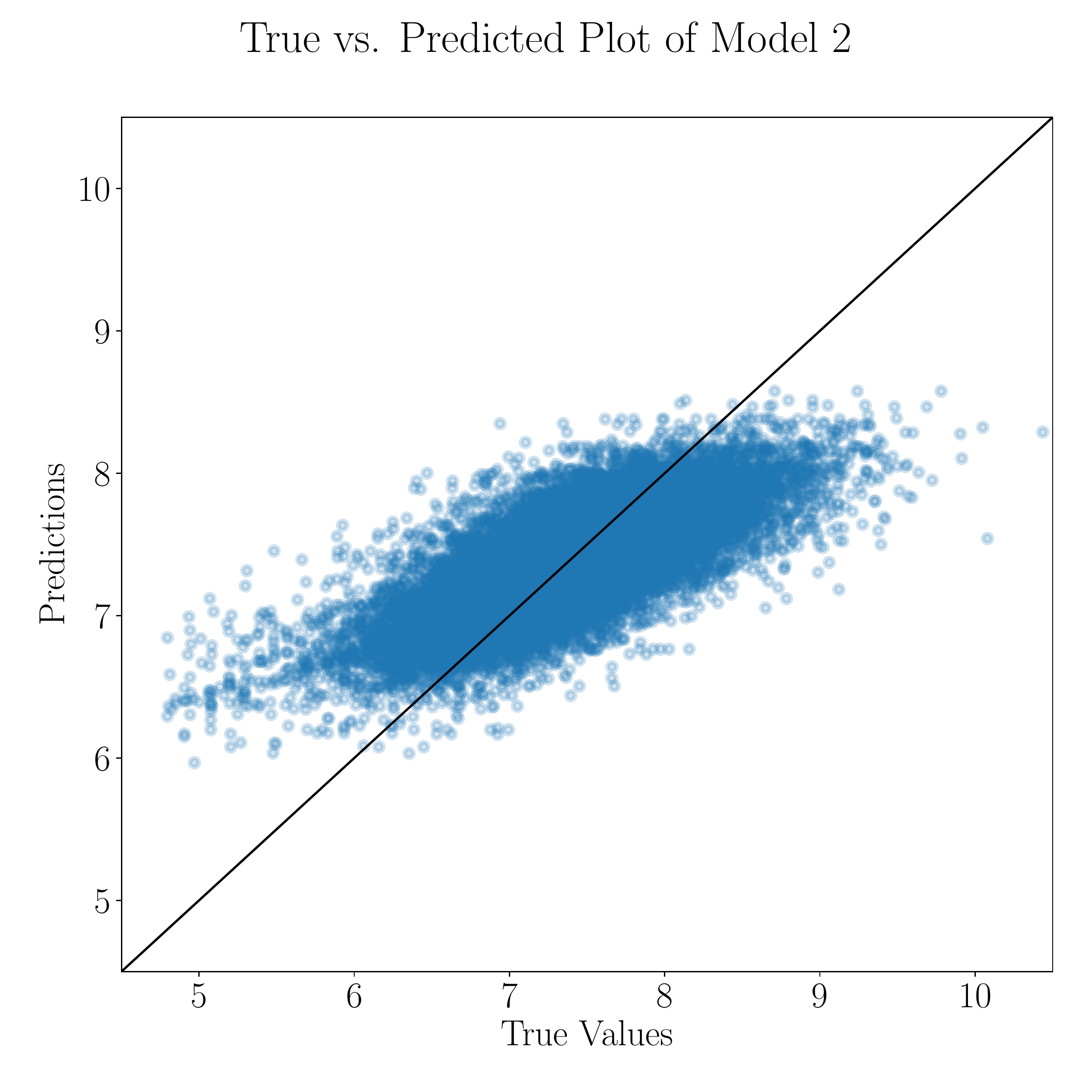}
    \caption{True values of the mean glance duration plotted against the predictions of Model 2 (linear mixed-effects model). All values are given on a logarithmic scale. Marginal $R^2 = 0.091$, Conditional $R^2 = 0.441$}
    \label{fig:TruePredictedPlot}
\end{figure*}

\end{document}